\documentclass[aps,pre,10pt]{revtex4-2}
\pdfoutput=1
\usepackage{graphicx}
\usepackage{amsmath}
\usepackage[pdfencoding=auto]{hyperref}
\newcommand*\mathinhead[2]{\texorpdfstring{$\boldsymbol{#1}$}{#2}}
\usepackage{wrapfig, blindtext}
\usepackage{subfigure}
\usepackage{amssymb}
\usepackage{xparse}
\ExplSyntaxOn
\AtBeginDocument
 {\linespread{ \fp_eval:n { 12pt/\baselineskip } }
  \selectfont}
\ExplSyntaxOff
\usepackage{tikz}
\usetikzlibrary{arrows,automata}
\usetikzlibrary{shapes,arrows}
\usetikzlibrary{intersections}
\usetikzlibrary{matrix,backgrounds}

\newcommand{\be}{\begin{equation}}
\newcommand{\ee}{\end{equation}}
\newcommand{\bea}{\begin{eqnarray}}
\newcommand{\eea}{\end{eqnarray}}
\renewcommand \thesubsection{\Roman{section}.\Alph{subsection}}
\date{\today}
\begin{document}
\title{Quantum heat engines with complex working media, complete Otto 
cycles and heuristics}
 \author{Venu Mehta and Ramandeep S. Johal} 
 \email[e-mail: ]{rsjohal@iisermohali.ac.in}
 \affiliation{Indian Institute of Science Education and 
 Research Mohali, Department of Physical Sciences, 
 Sector 81, S.A.S. Nagar, Manauli PO 140306, Punjab, India}
\begin{abstract}
Quantum thermal machines make use of 
non-classical thermodynamic resources, one of which is interactions
between elements of the quantum working medium. In this paper, we examine the 
performance of a quasi-static quantum Otto engine based on two spins of 
arbitrary magnitudes
subject to an external magnetic field and coupled via an isotropic Heisenberg 
exchange interaction. It has been earlier shown that the said
interaction provides an enhancement of cycle efficiency for two spin-1/2 
particles, with an upper bound
which is tighter than the Carnot efficiency. However, the necessary 
conditions governing engine performance and the relevant upper bound for
efficiency are unknown for the general case of arbitrary spin magnitudes.
Analyzing extreme-case scenarios, we formulate heuristics to 
infer the necessary conditions for an engine with uncoupled as 
well as coupled spins model. These conditions lead us to a connection 
between 
performance of quantum heat engines and the notion of majorization. 
Further, the study of complete Otto cycles inherent in the average cycle
also yields interesting insights into the average performance.
\end{abstract}
\maketitle
\section{Introduction}

Thermodynamics originated as an empirical study of steam engines, 
which blossomed into 
a framework of exceptional generality and simplicity.
Quantum thermodynamics is an emerging research field
that aims to extend classical thermodynamics and
statistical physics into the quantum realm---offering 
new challenges and opportunities in the wake of
a host of non-classical features.
A dominant interest is to understand energy-conversion processes 
at length scales and temperatures
where quantum effects become imperative. 
Inspired by our enhanced
capabilities towards nanoscale design and control, 
this endeavour is being pursued
by scientists from diverse backgrounds, such as statistical physics,
quantum information, quantum optics, many-body physics and so on.

To lay foundations for technological breakthroughs, a variety 
of fundamental questions are being addressed---ranging from issues
of thermalisation of quantum systems to examining the validity 
 of thermodynamic concepts, such as definitions of 
work, heat,  
efficiency and power at the nanoscale. The accord between 
quantum mechanics and thermodynamics is yet to fully unfold
\cite{gemmer2009quantum,kosloff2013quantum,binder2018thermodynamics}.
Its fundamental implications have inspired
numerous proposals for thermal machines based on quantum working media 
 \cite{Quan2007,AJM2008,thomas2011coupled,esposito2012stochastically,
kolavr2012quantum,levy2012quantum,PhysRevA.98.042102,AgarwalChaturvedi2013,
correa2013performance,del2014more,gelbwaser2013minimal,
venturelli2013minimal,PhysRevE.91.062137,Ou_2016,mehta2017quantum,Erdman2017,
Campo2017,PhysRevE.95.032111,agarwalla2017quantum,niedenzu2018quantum,PhysRevA.95.053870,XU2020135201,turkpencce2019coupled,ccakmak2019spin,xu2020unruh,
de2020two,huang2020two,zhang2020optimization,lee2020finite,PhysRevE.98.052147,
hong2020quantum,dey2019quantum,Latune_2019,de2019efficiency,
park2019quantum,johnson2020holographic,abah2020shortcut, VSJohal2020, 
myers2020bosons,PhysRevE.102.012138,makarov2020quantum,PhysRevE.101.012116,
PhysRevResearch.3.023078,PhysRevA.102.012217,PhysRevA.99.022129,
PhysRevE.103.032130}. Two major issues which are addressed 
in such proposals, are: What are the performance bounds of heat engines
working in quantum regime and what are the thermodynamic properties
of these quantum systems which control these bounds?
The performance analysis of various quantum analogues of classical
heat engines serve as test bed to study different extensions of thermodynamic
ideas in the
quantum world. With  the  recent  development  of  quantum  information
technology \cite{adesso2014continuous,scappucci2020germanium,goold2016role,
liu20192d}  and 
a  number  of  interesting  results,  the  study  of quantum heat engines 
(QHEs) has drawn much interest. In fact, the past few years
witnessed conducive studies exploring how the quantum statistics,
discreteness of energy levels, quantum adiabaticity, quantum coherence,
quantum measurement and entanglement affect the operation of heat
engines and cycles in various experimental set-ups including trapped ions, 
transmon qubits and more \cite{rossnagel2016single,maslennikov2019quantum,
ono2020analog,cimini2020experimental,benenti2017fundamental,
gelbwaser2015thermodynamics,rossnagel2014nanoscale,gelbwaser2018single,
zheng2015quantum,brantut2013thermoelectric,roulet2017autonomous,cherubim2019non,
peterson2019experimental,van2020single, ronzani2018, 
klatzow2019experimental,von2019spin,hicks1993effect,mahan1996best,
hicks1993thermoelectric,hartmann2015voltage,thierschmann2015three,
jaliel2019experimental,prance2009electronic,bloch2012quantum,cirac2012goals,
blatt2012quantum,ciani2019hamiltonian,bouton2021quantum,
solfanelli2021experimental}.

Finite time thermodynamic cycles 
\cite{singh2018low,Pozas_Kerstjens_2018,deffner2018efficiency,
camati2019coherence,mukherjee2020universal,hong2020quantum,lee2020finite,
chen2020power,beau2016scaling,wang2019finite,chand2021finite,denzler2020power,
Alecce_2015,e22101066,das2020quantum} and the study of open quantum systems 
\cite{pollock2018non,ingold2009specific,butanas2020dynamics,sone2020quantum,
latune2019energetic,rivas2020strong,PhysRevE.97.062108,santos2021maximally}
have drawn significant attention in the recent years. These studies aim
to arrive at more practical estimates of the performance measures for
these machines. However, the importance of quasi-static models of QHEs
lies in the fact that they provide a benchmark against which we can 
compare the behavior of finite time or more realistic models of heat
engines. A variety of quantum working substances have been
used to model these QHEs.
Amongst  these,  the  study  of simple, coupled  quantum  systems 
\cite{thomas2011coupled,huang2013quantum,e21111131,Huang2014,
altintas2014quantum,altintas2015general,
ivanchenko2015quantum,mehta2017quantum,zhao2017entangled,huang2020two,
alet2020entanglement,zhang2020optimization, Jonathan2020, PhysRevE.87.012144}
can yield important insights into the role of quantum interactions
in enhancing the performance of model thermal machines.
In particular, an upper bound ($\eta_{\rm ub}$)
for quantum Otto efficiency using coupled spin-$1/2$ particles
has been obtained which 
is tighter than the Carnot bound ($\eta_{\rm C}$) 
\cite{thomas2011coupled,mehta2017quantum}.
However, this upper bound seems to be violated
for coupled spins with higher magnitudes \cite{altintas2015general}.

It is apparent that as the quantum working medium becomes complex--as
for a many-body system or when the energy spectrum is non-trivial, 
an exact analysis may become intractable. 
This is especially true when
the working medium is neither a few-particles system with 
a simple energy spectrum nor a medium close to thermodynamic
limit where some scaling law may aid in mathematical simplicity 
\cite{Campisi2016}.
Thus, to target this intermediate regime, it seems useful  
to formulate heuristics.
The latter are of significance in various disciplines
such as cognitive science, behavioral economics and
computer science, to name a few. Broadly speaking,
a heuristic is a rule of thumb providing insights 
into the behavior of a system in the face of complexity or
uncertainty \cite{Polya1971,Simon1958Heuristic,Gigebook2011}. 
 The solution suggested by a heuristic may not be optimal or  
may simply be an approximate solution. However, the value of 
a heuristic lies in providing a shortcut method 
that requires a simpler
analysis, thus trading accuracy and completeness for 
speed. 

Based on this understanding, we analyze 
 complete Otto cycles (COCs) to characterize the performance
of our engine. In a COC, the working medium starts and 
ends in the same state. In general, during thermal
interaction with reservoirs, the transitions in 
the system are not deterministic. We show that 
COCs which follow the second
law under a certain operation (say as an engine), 
also yield conditions to analyze the global or average performance
of the machine.

In this paper, we carefully make use of the information 
discerned from the energy-level structure of the working medium, 
as well as general relations between the canonical probabilities
arising from interactions with heat reservoirs.
The worst-case or best-case scenarios (WCS/BCS) under a 
given situation are employed to infer
necessary conditions for an Otto engine. Thus, we are
able to derive positive work condition and establish consistency
with the second law of thermodynamics. We also infer
an upper bound for the efficiency of the Otto cycle
setting new benchmarks for Otto efficiency that is
tighter than Carnot limit.

The paper is organized as follows. In Section II, we introduce our model
of two coupled spins ($s_{1}, s_{2}$) 
as the working substance of the Quantum Otto engine.
In Section II.A, various stages of the heat cycle are described
and positive work condition for the uncoupled model is discussed. 
The proof for the same is sketched in Appendix A. 
In Section III, the spins are coupled 
and we find the coupling range in which positive work
extraction  is ensured (proofs are sketched
in Appendices B and C) which is related to the notion
of majorization in Section III.A and
further used to order the system's
energy levels for $J \ne 0$ in Section III.B.  
In Section IV conditions for maximal
enhancement of coupled system's efficiency over the uncoupled
model are discussed. An upper bound to engine's  efficiency 
is also calculated in the considered domain of coupling. 
A detailed proof for the positive entropy production for
the coupled system is sketched in Appendix D. In Section V, 
an analysis is carried out using the notion of complete Otto 
cycles. Finally, we discuss the results of our analysis in
Section VI.
\section{Quantum Otto cycle}
The working substance consists of two spins with
arbitrary magnitudes, $s_1$ and $s_2$, coupled by 1-D isotropic
Heisenberg exchange interaction, in the presence of an
externally applied magnetic field of magnitude $B$ along $z$-axis. 
The system Hamiltonian in the first Stage of the cycle 
can be written as:
\begin{equation}
\mathcal{H}_{1} \equiv H_{1}+H_{int}=
2B_{1} \left(s_{1}^{(z)} \otimes I  + I \otimes s_{2}^{(z)} \right)
+8J\vec{s}_{1}.\vec{s}_{2} 
\label{Ham}
\end{equation}
\noindent where $J>0$ is the strength of the 
anti-ferromagnetic coupling.
$\vec{s}_{1} \equiv \{ s_{1}^{(x)}, s_{1}^{(y)}, s_{1}^{(z)} \}$,
$\vec{s}_{2} \equiv \{ s_{2}^{(x)}, s_{2}^{(y)}, s_{2}^{(z)} \}$
are the spin operators for the first and the second spin 
respectively. $H_{int}$ is the interaction Hamiltonian
and $H_{1}$ is the free Hamiltonian.
We have taken Bohr magneton $\mu_{\rm B} =1$ 
and the gyromagnetic ratio for both spins has been taken to be 2 
\cite{Ferrara1992}.

Let
\textit{n=$\left(2s_{1} +1\right)\left(2s_{2} +1\right)$}
be the total number of energy levels with ${\left| \psi _{k} \right\rangle}$
as the corresponding energy eigenstates. When the system 
is in thermodynamic equilibrium with a heat bath at temperature $T$,
the density matrix $\rho_{1}$ for the working substance 
can be written as:
\begin{equation}
\rho_{1} =\sum _{k=1}^{n}P_{k}  {\left| \psi _{k}
\right\rangle} {\left\langle \psi _{k}  \right|}, 
\label{rho}
\end{equation} 
where $P_{k} =e^{-{E_{k}}/{T}}/{Z}$
are the occupation probabilities of the energy
levels and $Z=\sum _{k} e^{-{E_{k}}/{T}}$ 
is the partition function for the system.
We have put the Boltzmann constant $k_{\rm B}$ equal to unity.

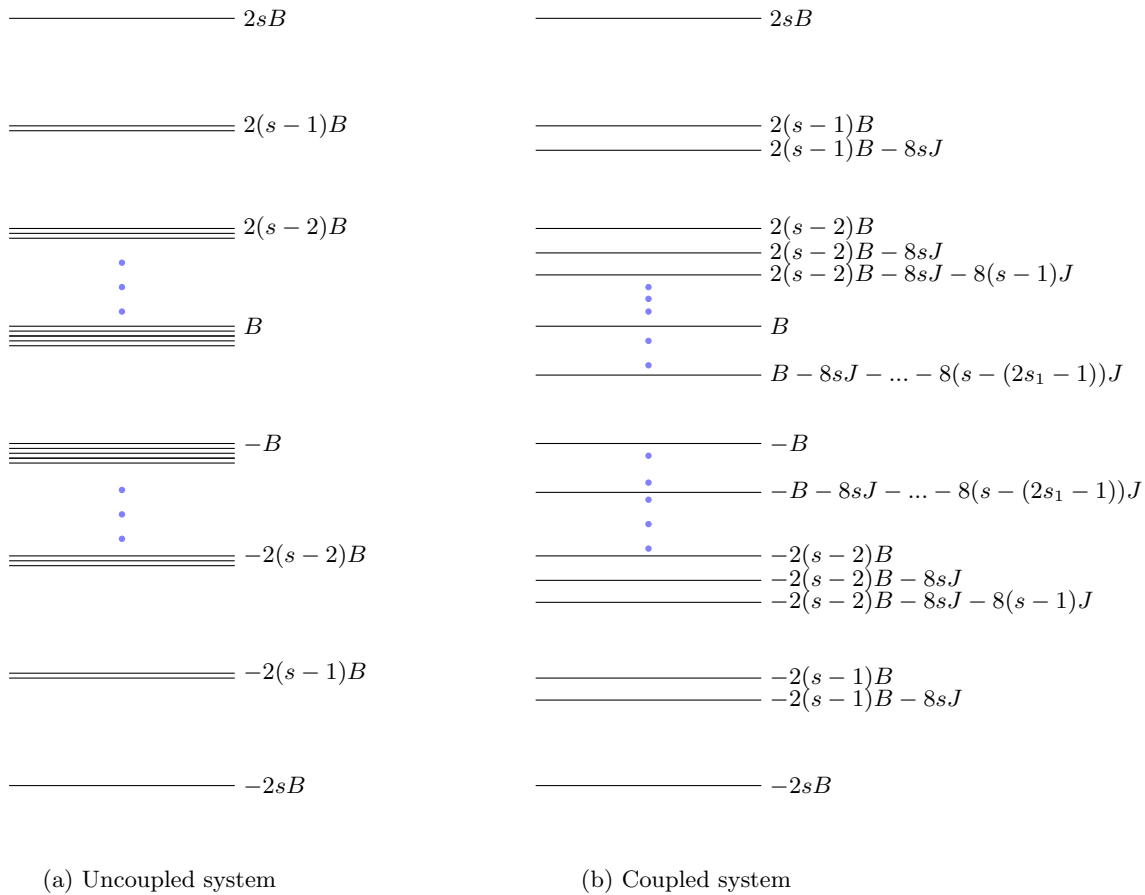
\begin{figure}
\begin{tikzpicture}
\scalebox{10}{5}
 
 \draw[] (2cm,0em) -- (5cm,0em) node[right] {$-2sB$};
 \draw (4,-1)  node[above]
 {} node[below] {(a) Uncoupled system}
 (5,-1);
 \draw (11,-1)  node[above]
 {} node[below] {(b) Coupled system}
 (12,-1);
 \draw (9cm,0em) -- (12cm,0em) node[right]
{$-2sB$};
\draw (9cm,3.5em) -- (12cm,3.5em) node[right]
{$-2(s-1)B-8sJ$};

\draw (9cm,4.4em) -- (12cm,4.4em) node[right]
{$-2(s-1)B$};

\draw (9cm,7.5em) -- (12cm,7.5em) node[right]
{$-2(s-2)B-8sJ-8(s-1)J$};

\draw (9cm,8.4em) -- (12cm,8.4em) node[right] 
{$-2(s-2)B-8sJ$};

\draw (9cm,9.4em) -- (12cm,9.4em) node[right]
{$-2(s-2)B$};

\filldraw [blue!50] (10.5cm,9.7em) circle (1pt);
\filldraw [blue!50] (10.5cm,10.7em) circle (1pt);
\filldraw [blue!50] (10.5cm,11.7em) circle (1pt);
\draw (9cm,12em) -- (12cm,12em) node[right]
{$-B-8sJ-...-8(s-\left(2s_{1} -1\right))J$} ;
\filldraw [blue!50] (10.5cm,12.4em) circle (1pt);
\filldraw [blue!50] (10.5cm,13.5em) circle (1pt);

\draw  (9cm,14em) -- (12cm,14em) node[right]
{$-B$};

\draw (9cm,16.8em) -- (12cm,16.8em) node[right]
{$B-8sJ-...-8(s-\left(2s_{1} -1\right))J$};

\filldraw [blue!50] (10.5cm,17.2em) circle (1pt);
\filldraw [blue!50] (10.5cm,18.2em) circle (1pt);

\draw (9cm,18.8em) -- (12cm,18.8em) node[right]
{$B$};
\filldraw [blue!50] (10.5cm,19.4em) circle (1pt);
\filldraw [blue!50] (10.5cm,20.4em) circle (1pt);
\filldraw [blue!50] (10.5cm,19.92em) circle (1pt);
\draw (9cm,20.9em) -- (12cm,20.9em) node[right]
{$2(s-2)B-8sJ-8(s-1)J$};
\draw (9cm,21.8em) -- (12cm,21.8em) node[right]
{$2(s-2)B-8sJ$};
\draw (9cm,22.8em) -- (12cm,22.8em) node[right] 
{$2(s-2)B$};
\draw (9cm,26em) -- (12cm,26em) node[right]
{$2(s-1)B-8sJ$};
\draw (9cm,27em) -- (12cm,27em) node[right] 
{$2(s-1)B$};
\draw (9cm,31.4em) -- (12cm,31.4em) node[right]
{$2sB$};
    \draw[] (2cm,4.4em) -- (5cm,4.4em);
        \draw[] (2cm,4.6em) -- (5cm,4.6em) node[right]{$-2(s-1)B$};
    \draw[] (2cm,9em) -- (5cm,9em)  ;
   \draw[] (2cm,9.2em) -- (5cm,9.2em);
   \draw[] (2cm,9.4em) -- (5cm,9.4em) node[right] {$-2(s-2)B$};
   \filldraw [blue!50] (3.5cm,10.1em) circle (1pt);
   \filldraw [blue!50] (3.5cm,11.1em) circle (1pt);
   \filldraw [blue!50] (3.5cm,12.1em) circle (1pt);
   \draw[] (2cm,13.2em) -- (5cm,13.2em) node[right] {} ;
   \draw[] (2cm,13.4em) -- (5cm,13.4em) node[right] {} ;
   \draw[] (2cm,13.4em) -- (5cm,13.4em)  ;
   \draw[] (2cm,13.6em) -- (5cm,13.6em) node[right] {} ;
   \draw[] (2cm,13.8em) -- (5cm,13.8em) node[right] {$$} ;
   
   \draw[] (2cm,14em) -- (5cm,14em) node[right] {$-B$} ;
     \draw[] (2cm,18.4em) -- (5cm,18.4em) node[right] {};
\draw[] (2cm,18em) -- (5cm,18em) node[right] {};
       \draw[] (2cm,18.2em) -- (5cm,18.2em);
        \draw[] (2cm,18.4em) -- (5cm,18.4em) node[right] {};
         \draw[] (2cm,18.6em) -- (5cm,18.6em) node[right] {$$};
         
          \draw[] (2cm,18.8em) -- (5cm,18.8em)  node[right] {$B$};
           
        \filldraw [blue!50] (3.5cm,19.4em) circle (1pt);
   \filldraw [blue!50] (3.5cm,20.4em) circle (1pt);
   \filldraw [blue!50] (3.5cm,21.4em) circle (1pt);
   
   \draw[] (2cm,22.4em) -- (5cm,22.4em) node[right] {};
    \draw[] (2cm,22.6em) -- (5cm,22.6em) node[right] {};
     \draw[] (2cm,22.8em) -- (5cm,22.8em) node[right] {$2(s-2)B$};
    \draw[] (2cm,26.8em) -- (5cm,26.8em) node[right] {};
      \draw[] (2cm,27em) -- (5cm,27em) node[right] {$2(s-1)B$};
      \draw[] (2cm,31.4em) -- (5cm,31.4em) node[right] {$2sB$};
\end{tikzpicture}\\
\caption{The above figure shows the energy levels of the two-spins 
($s_{1},s_{2}$) system for (a) $J=0$ and (b) $J>0$. The degeneracy
in the energy levels is lifted as the interaction is switched on ($J\ne0$).  
Here $s_{1}<s_{2}$ and $s=s_{1}+s_{2}$. }
\label{fig:1}
\end{figure}

Let us consider the case where one spin is an 
integer and other is a half integer. Some examples of such spin combinations 
are
\noindent $\left(\frac{3}{2},2\right),
\left(\frac{1}{2},2\right),\left(\frac{5}{2}, 4\right)$.
The energy eigenvalues of the Hamiltonian $\mathcal{H}$ for a general
($s_{1},s_{2}$) coupling are shown in Fig. \ref{fig:1}.
It is to be noted that a term $8s_{1}s_{2}J$ common 
in all the eigenvalues has been neglected as
the physical properties of the system would be independent of it.
\noindent The ordering of these energy levels 
would depend upon the conditions on the parameters
which the positive work condition for the system would
provide, which will be discussed in the coming sections.
\subsection{The heat cycle}
The four stages constituting the Otto cycle are as follows.
\\
\noindent \textit{Stage 1}: The system is at thermal equilibrium 
 with a heat reservoir at temperature $T_{1}$ with energy $e_{k}$
 whose occupation probabilities are $p_{k}$ and the corresponding
 density matrix is $\rho_{1}$ (here we are
 considering two non interacting spins 
 with energy eigenvalues denoted by $e_{k}$
 and occupation probabilities by $p_{k}$). 
\begin{table}
\centering
\begin{tabular}{p{6cm} p{5cm}} 
 \hline
 $k$ & $e_{k}$ \\ 
 \hline\hline
$1$ & $-2sB$  \\ 
 \hline
$2,3$ & $-2(s-1)B$  \\
 \hline
 $4,5,6$ & $-2(s-2)B$  \\
 \hline
 . & .\\
  . & .\\
  \hline
  $(n/2-2s_{1}),...,n/2$ & $-2(s-r)B$ \\
  \hline
 $(n/2+1),...,(n/2+2s_{1}+1)$ &  $2(s-r)B$  \\
  \hline
  . & .\\
  . & .\\
  \hline
 $(n-5),(n-4),(n-3)$ & $2(s-2)B$  \\ 
 \hline
$(n-2),(n-1)$ & $2(s-1)B$  \\
 \hline
 $n$ & $2sB$  \\
 \hline
 \hline
\end{tabular}
\caption{Levels indicating degeneracy and energy eigenvalues 
($e_{k}$) for two uncoupled spins. Here, $s_{1} < s_{2}$ with 
$s =s_1+s_2$ and $r \equiv s-1/2$.}
\label{table:1}
 \end{table} 
\par \noindent 
\textit{Stage 2}: The system undergoes 
a quantum adiabatic process after it is isolated
from the hot bath and the magnetic field is changed
from $B_{1}$ to a smaller value $B_{2}$. Here,
the quantum adiabatic theorem is assumed to hold according to
which the process should be slow enough so that 
no transitions are induced as 
the energy levels change from $e_{k} $
to $e_{k}^{'}$.
\\
\noindent \textit{Stage 3}:
Here the system 
is brought in contact with a cold bath at
temperature $T_{2} (<T_{1})$.
The energy eigenvalues remain at $e_{k}^{'}$
and the occupation probabilities change
from $p_{k}$ to $p_{k}^{'}$ with the 
external magnetic field at $B=B_{2}$ 
and the density matrix of the system 
is $\rho_{2}$. \\
\noindent \textit{Stage 4}:
The system is detached 
from the cold bath and the magnetic field is
changed from $B_{2}$ to $B_{1}$  with occupation
probabilities remaining unchanged at $p_{k}^{'}$
and energy eigenvalues change back from $e_{k}^{'}$ 
to $e_{k}$ such that only work is performed on the 
system during this step.
\noindent Finally, the system is attached to the
hot bath again and the cycle is completed such
that the average heat absorbed is 
 $q_{1,av}={\rm Tr}[H_{1}\Delta\rho_{}]$,
and the net work performed per cycle is
$w_{av}={\rm Tr}[(H_{1}-H_{2}) \Delta\rho_{}]$.
Here, ${\rm Tr}[\cdot]$ denotes the trace operation, 
and $\Delta\rho_{}=\rho_{1}-\rho_{2}$. In this paper, 
we consider the free Hamiltonian 
of the form $H_{i} \equiv 2B_{i}h_{0}$ ($i=1,2$), 
where $h_0$ is an operator. 
We now have, $w_{av}=2(B_{1}-B_{2}){\rm Tr}[h_{0} \Delta\rho_{}]$,
and therefore, the efficiency in the absence of interaction is
\begin{equation}
\eta_{0}=1-\frac{B_{2}}{B_{1}}.
 \label{eta0}
\end{equation}
\noindent Let us first discuss about
the positive work condition when $s_{1}$ 
and $s_{2}$ are non-interacting. The energy
eigenvalues ($e_{k}$) of the free Hamiltonian, written in 
the order of increasing energy (if one spin
is integer and the other is half integer)
are listed in Table \ref{table:1} and as 
can be seen many energy levels for the 
non-interacting system are degenerate.
There is only one level with energy 
proportional to $-s$ as well as  $s$, two levels
with energy proportional to $-(s-1)$ as well as
$(s-1)$ and so on, the proportionality constant
always being $2B$. Therefore, 
denoting the degeneracy by $"g"$ we have the
following from Table \ref{table:1},
\begin{equation}
g_{\lvert{s} \rvert}=1,\quad g_{|s-1|}=2,
\quad g_{|s-2|}=3 ,...,g_{|s-r|}=2s_{1}+1
\label{deg}
\end{equation}
such that the total number of energy levels are, 
$n=(2s_{1}+1)(2s_{2}+1)=2(g_{|s|}+g_{|s-1|}+g_{|s-2|}
+...+g_{|s-r|})$.

The \textit{Stage 1} occupation probabilities are written as, 
$p_{k} = {e^{- e_{k}/T_1}}/{z_{1}}$, where
$z_{1}={\sum_{k=1}^{n}e^{- e_{k}/T_1}}$ is the partition function 
of the system which can be expressed as follows.
\begin{equation}
z_{1} =2\sum_{l=1}^{s+1/2} g_{|s-l+1|}.
\cosh{\left[2(s-l+1)B_{1} /T_{1} \right] }.
\label{z1}
\end{equation}
\noindent The average heat exchanged with the hot reservoir is
\begin{equation}
q_{1,av} =\sum_{k=1}^{n} e_{k}
\left(p_{k} -p_{k}^{'} \right) =2B_{1}v,
\label{qh}
\end{equation}
where the primed probabilities are
tabulated at $T=T_{2} $ and $B=B_{2}$.
The average heat exchanged with the cold bath is
\be
q_{2,av}=\sum_{k=1}^{n} e_{k}^{'} \left(p_{k}^{}
-p_{k}^{'} \right)
=2B_{2} v,
\ee
\noindent so that the work done on average is:
\begin{equation}
    w_{av}=q_{1,av} -q_{2,av} =2(B_{1} -B_{2} )v.
    \label{work}
\end{equation}
The explicit expression of $v$ is given by Eq. (\ref{proof}).
 Since $B_{1} >B_{2}$ is assumed, the system works as an 
engine on average, iff 
   $v>0$.
We prove in Appendix A 
that the condition required to satisfy $v>0$ is 
\begin{equation}
 \frac{B_{2}}{T_{2}}>\frac{B_{1}}{T_{1}}, \;\; {\rm or} \;\; B_2 > B_1 \theta, 
\label{pwc}   
\end{equation} 
where $\theta = T_2/T_1$.
Further, as proved in Appendix A, Eq. (\ref{pwc}) implies  $z_{2}>z_{1}$
as well as
\be
 p_{1}^{'} > p_{1}, \quad {\rm and} \quad p_{n}^{'}< p_{n},
 \label{p1p}
\ee 
From the above conditions, we can make the 
following inferences.
Positive work extraction is favoured when 
the occupancy of ground (top) level is
more (less) at the cold bath than at the
hot bath which suggests that heat is absorbed
at the hot bath, decreasing (increasing)
the occupancy of the ground (top) level, while heat
is released at the cold bath, thus increasing 
(decreasing) the occupancy of the ground (top)
level. 
 
 Since the
 working medium returns to its initial state (restoring 
 the Hamiltonian as well as coming to be in equilibrium
 with the hot reservoir), the net change in 
 entropy $\Delta S_{0,av}$  is due to 
 the entropy changes only in the heat baths.
 The decrease in the entropy of the hot bath is
 $-{q_{1,av}}/{T_{1}}$ and increase in entropy
 of the cold bath is ${q_{2,av}}/{T_{2}}$. 
 So, the net entropy change in one cycle is,
 \begin{equation}
 \Delta S_{0,av}=-\frac{q_{1,av}}
 {T_{1}}+\frac{q_{2,av}}{T_{2}} = \left(-\frac{B_{1}}{T_{1}}+
 \frac{B_{2}}{T_{2}}\right)v.
 \label{ds0}
 \end{equation}
We have seen that $w_{av}>0$ or $v>0$
requires Eq. (\ref{pwc}) to hold.
Under these conditions, it follows that $\Delta S_{0,av} >0$
and so the consistency with the second law is 
established at the level of average performance as an engine. 
Similarly, we observe that the  
 efficiency satisfies:
 $\eta_{0}<
1-{T_{2}}/{T_{1}}=\eta_{C}$.

\section{The coupled model}
Let us now couple the two spins, with $J>0$ being the
anti-ferromagnetic coupling strength. The corresponding energy 
eigenvalues are shown in Fig. \ref{fig:1}b,
where the ordering of the eigenvalues can be considered  when the coupling 
parameter $J$ is small.
Also, as the coupling is switched on, the degeneracy of
the previously degenerate levels is now lifted. Let us express an energy 
eigenvalue of the coupled system as: $E_{k}=m_{1} B-8m_{2} J$,
 where $m_{1}=-2s,...,+2s$ and 
 $m_{2}$ can only take positive values 
 including zero, as shown in Table \ref{table:5} in Appendix C.
 The values $m_1$ and $m_2$ depend on the index $k$, but we have
 omitted it here for brevity of notation.
 \\ 
Now,  the average heat
absorbed from the hot bath ($Q_{1,av}$), 
the heat rejected to the cold bath ($Q_{2,av}$)
and the average work done in one 
cycle, 
$W_{av} = Q_{1,av} - Q_{2,av}$, are given as
 \bea
 Q_{1,av} &=& 2B_{1}X+8JY, \nonumber \\ 
 Q_{2,av} &=& 2B_{2}X+8JY, \label{Q12W} \\
 W_{av} &=& 2(B_{1}-B_{2})X, \nonumber
\eea 
where
\begin{equation}
X= \cfrac{1}{2}\sum_{k=1}^{n} m_{1} (P_{k}^{}-
P_{k}^{'}), \quad Y= \sum_{k=2}^{n-2} 
m_{2} (P_{k}^{'}-P_{k}).
\label{xy}
\end{equation} 
The spin dependent factors $m_{1}$ and $m_{2}$ are obtained from the 
expressions of the equilibrium occupation probabilities of the energy levels 
$E_{k}$ (shown in Fig. \ref{fig:1}), which in general are written as, 
\begin{equation}
    P_{k} = \cfrac{e^{-m_{1} B_{1}/T_{1} +8m_{2} J/T_{1}}}{Z_{1}}.
    \label{Pk}
\end{equation} For explicit expressions of $P_{k}$, refer to 
Table \ref{table:2} in Appendix B.
$Z_{1}$ is the \textit{Stage 1} partition 
function of the system whose expression may be rewritten as,
\begin{equation}
Z_{1}= \left\{ \begin{array}{l} {\cal Z}_{1} +2\cosh 
\left[2(s-1)B_{1}/T_{1}\right].e^{8sJ/T_{1}}+\\ 
2\cosh \left[2(s-2)B_{1}/T_{1}\right].
\left(e^{8sJ/T_{1}}+e^{8(s-1)J/T_{1}}\right)+...+ \\ 
2\cosh \left[2(s-r)B_{1}/T_{1}\right].
\left(e^{8sJ/T_{1}}+e^{8(s-1)J/T_{1}}+...
+e^{8(s-(2s_{1}-1))J/T_{1}}\right),
\end{array} \right. 
 \label{Z}
\end{equation}
where 
${\cal Z}_{1} \equiv 2\sum_{k=1}^{s+1/2}
\cosh{\left[2(s-k+1)B_{1} /T_{1}\right]}$.  
Similarly, we can define $P_{k}^{'}$, the canonical
probabilities due to cold bath, by replacing 
$B_1 \to B_2$ and $T_1 \to T_2$ in the above expressions
for $P_{k}^{}$.

For the proof of PWC for the coupled model (Appendix B),    
we show that for
the so-called worst case scenario (WCS), given by
\be
     P_{k}^{'}  < P_{k}^{}, \; k=2,3,...,n, \;{\rm and}\; P_{1}^{'} > P_{1},
    \label{WCS1}
\ee  
along with Eq. (\ref{pwc}), 
 it follows that $X>0$.
Consistent with Eqs. (\ref{WCS1}) and  (\ref{WCS2}), we then calculate 
the strictest condition on the allowed range of $J$
(Appendix C) which is given by
\begin{equation}
    0<J<\frac{B_{2} -B_{1} \theta }{4s\left(1-\theta \right)}
    \equiv J_{c}.
    \label{jc0}
\end{equation}
Therefore, we conclude that
$X>0$ or PWC is satisfied under Eqs (\ref{pwc}) and (\ref{jc0})
with the latter constituting the
sufficient condition for the coupled 
system to work as an engine.
\subsection{Majorization}
Majorization \cite{marshall1979inequalities} is a powerful mathematical concept
that defines a preorder on the vectors of real
numbers. It is particularly useful 
to compare two probability distributions. 
We will highlight its occurance in the context
of the working regime of our engine by comparing
the two equilibrium probability distributions. 

Now, for the uncoupled model, the relevant probability
distributions are the canonical probabilities $\{ p_k\}$
and $ \{ p_{k}^{'} \}$, which, at finite temperatures, are ordered as:
$p_{n}^{} < p_{n-1}^{} <\cdots < p_{1}^{}$ and
$p_{n}^{'} < p_{n-1}^{'} <\cdots < p_{1}^{'}$, respectively.
In Lemma 2 of Appendix A, we proved that Eq. (\ref{pwc}) is a necessary 
condition that ensures $w_{av}>0$, in the regime of the so-called
worst case scenario (WCS), given by 
\[
  p_{k}^{'} \le p_{k}^{}, \quad k=2,3,...,n \quad \mathrm{and} \quad p_{1}^{' }
  \ge p_{1}^{},
\]
 where the equality holds for $B_{2}/T_{2}=B_{1}/T_{1}$. Therefore,
 the above relations imply
\begin{align*}
 p_{n}^{'} &\le  p_{n}, \nonumber \\
 p_{n}^{'}+p_{n-1}^{'} &\le  p_{n}^{}+p_{n-1}^{}, \nonumber \\
   & \vdots  \tag{M} \label{major} \\
 \sum_{k=1}^{n-1} p_{k}^{'} &\le  \sum_{k=1}^{n-1} p_{k}, \nonumber \\
 \sum_{k=1}^{n} p_{k}^{'} &= \sum_{k=1}^{n} p_{k}. \nonumber
 \end{align*} 
The above set of conditions (M)
is summarised by stating that $ \{ p_{k}^{'} \}$ {\it majorizes}
$ \{ p_{k}^{} \}$, and denoted as 
   $ \{p_{k}\} \prec \{p_{k}^{'}\}$. As a powerful tool,
   majorization can be used to prove other results too.
   Intuitively, it indicates that the distribution $ \{ p_{k}^{} \}$
   is more mixed than $\{p_{k}^{'}\}$.
   Thus, as an important consequence,  
   $ \{p_{k}\} \prec \{p_{k}^{'}\}$ implies that 
   $S(p_{k}) \ge S(p_{k}^{'})$, where $S(p)$ is the Shannon
   entropy of the distribution $\{p\}$
   (proportional to the thermodynamic entropy
   of the working medium in equilibrium with a reservoir).
   In fact, this
is expected, since the flow of heat for the engine is
on the average from hot to cold. 
Then, along with heat, thermodynamic entropy is 
also lost to the cold reservoir. However, 
the condition of majorization is more general
than the above mentioned relation between the entropies.
   
Similarly for the coupled model, we have shown that Eqs. (\ref{pwc}) 
and (\ref{jc0}) ensure $W_{av}>0$ under the conditions:
  $P_{k}^{'} < P_{k}^{}, \forall \; k=2,3,...,n$ {and} $P_{1}^{'}>P_{1}^{}$. 
In general, we may write 
\be
 P_{k}^{'} \le P_{k}^{}; \quad k=2,3,...,n \quad \mathrm{and} 
 \quad  P_{1}^{'}\ge P_{1}^{}.
 \label{wcsP}
\ee 
Thus, for the coupled model too, we can
write down the set of conditions equivalent to Eq. (M),
and infer that $\{P_{k}\} \prec \{P_{k}^{'}\}$, 
which implies 
$S(P_{k}) \ge S(P_{k}^{'})$. In other words, if the 
\textit{Stage 3} equilibrium distribution  majorizes \textit{Stage 1} 
equilibrium  distribution, 
 then we have positive work extraction from the coupled system. 
 
\begin{figure}
  \subfigure[]{\includegraphics[width=6cm,height=6cm,
  keepaspectratio]{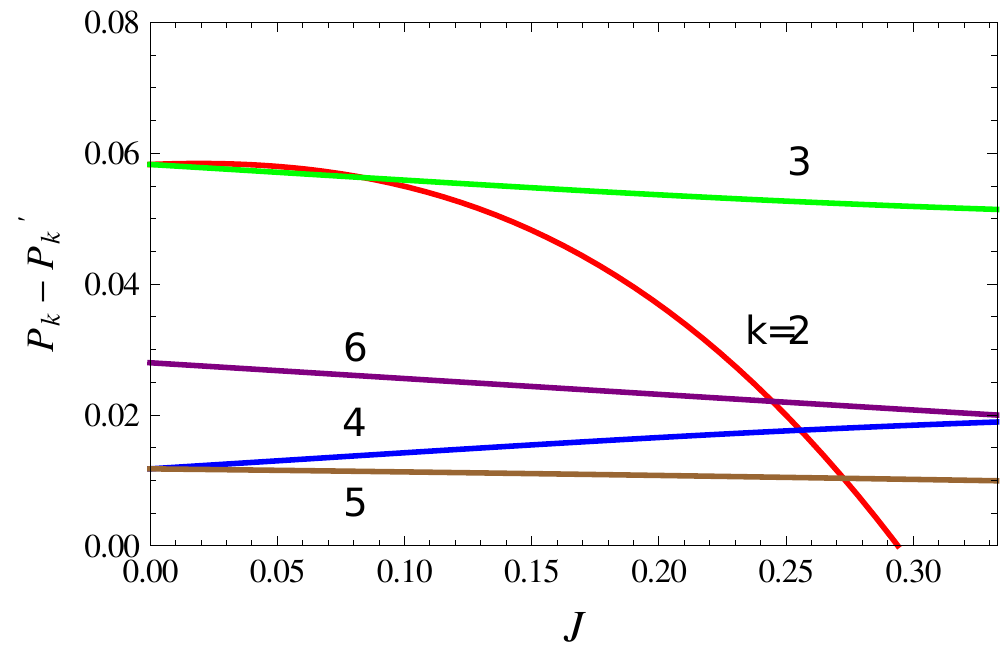}} 
    \subfigure[]{\includegraphics[width=6cm,height=6cm,
    keepaspectratio]{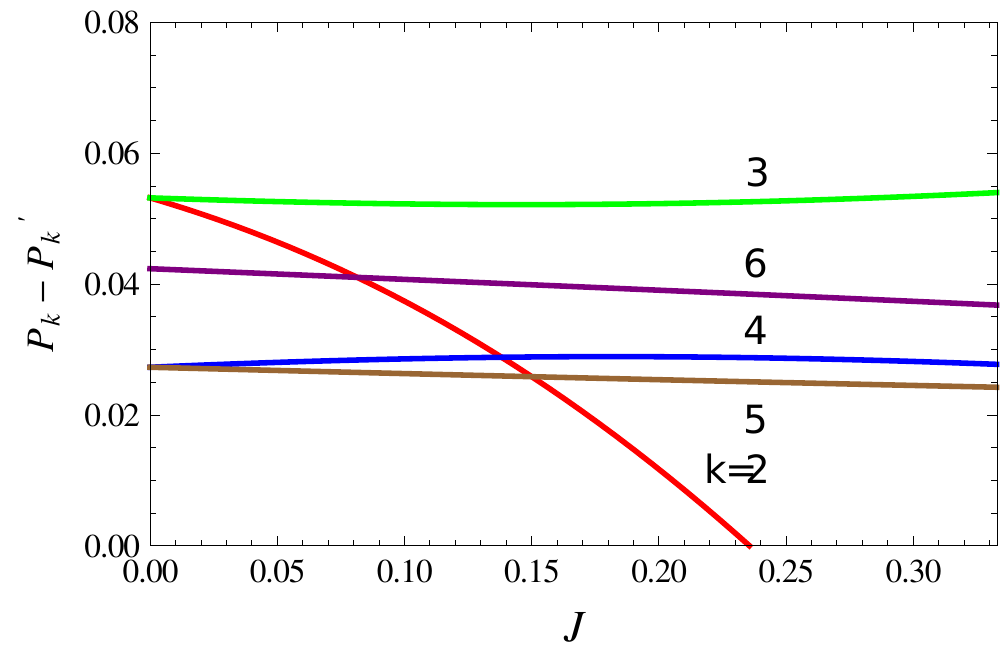}}
\caption{(a) Variation of $P_{k}^{}-P_{k}^{'}$ with the coupling factor 
$J$ for $(1/2,1)$ system,
with $k$ values ranging from 2 to 6.
 The parameters are set at $B_{1}=4,B_{2}=3$,
 with temperatures, a)
 $T_{1}=4,T_{2}=2$ and b) $T_{1}=6,T_{2}=3$. Here, $J_{c}=1/3$.
 The value of $J$ for which $P_{2}^{}-P_{2}^{'}$ (red curve) changes sign
 (from positive to negative)
approaches $J_c$ for lower temperatures (see also Fig. \ref{figmaj}).
 }
    \label{figpdiff}
\end{figure} 

\begin{figure}
 \subfigure[]{\includegraphics[width=6cm,height=6cm,
 keepaspectratio]{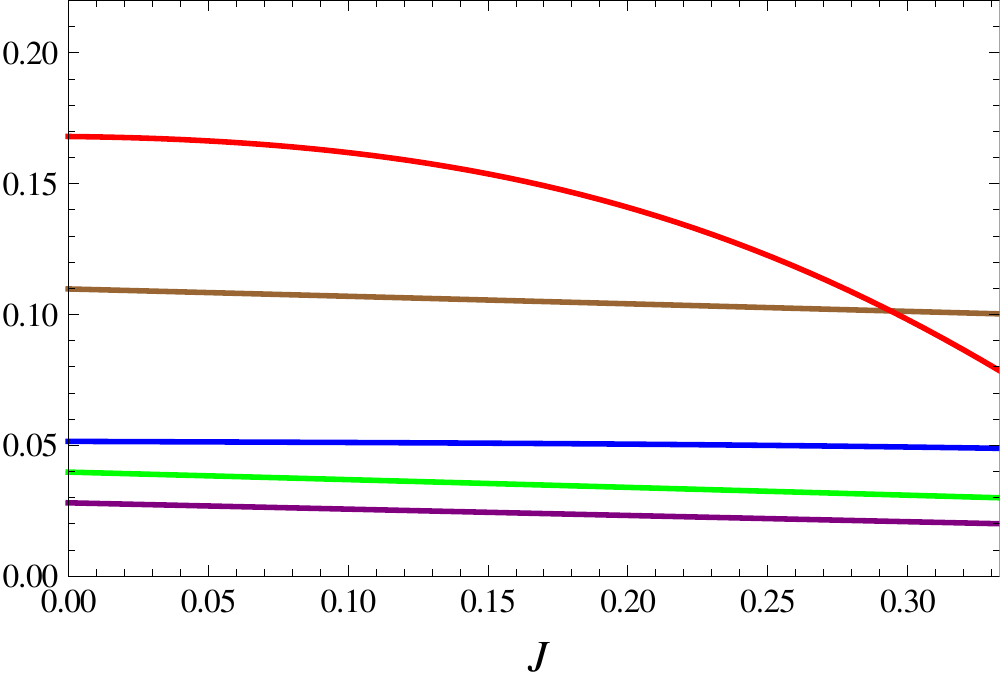}} 
    \subfigure[]{\includegraphics[width=6cm,height=6cm,
    keepaspectratio]{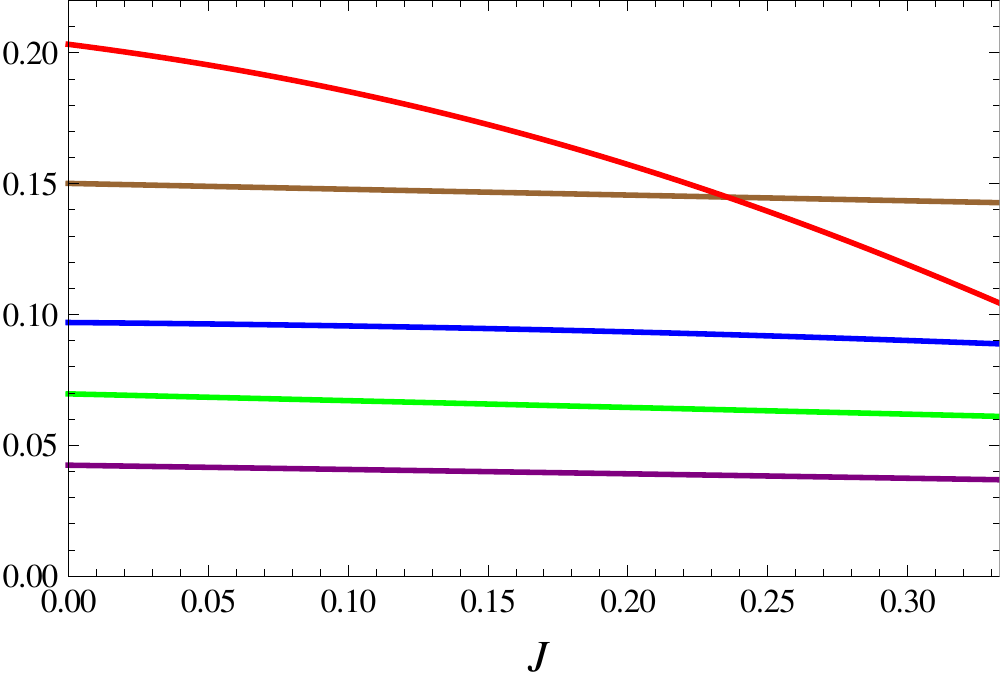}}
    \caption{(a) Majorization conditions shown by positivity of all
    quantities $P_{6}^{} - P_{6}^{'}$ (purple), $\sum_{k=5}^{6} P_{k}^{}
    - P_{k}^{'}$ (green),
    $\sum_{k=4}^{6} P_{k}^{} - P_{k}^{'}$ (blue), $\sum_{k=3}^{6} P_{k}^{}
    - P_{k}^{'}$ (brown),
    and $\sum_{k=2}^{6} P_{k}^{} - P_{k}^{'}$ (red) as function of 
the coupling strength $J$ for ($1/2,1$) system of Fig. \ref{figpdiff}. 
The point where the red curve intersects the lower curve is where 
$P_{2}^{'} = P_2$.
It is seen that for higher bath temperatures (for a given ratio $T_2/T_1$),
this point shifts to lower $J$ values.
}
    \label{figmaj}
\end{figure}
It is possible to find a range of parameter values which satisfy 
Eq. (\ref{wcsP}).
In Fig. \ref{figpdiff}, we show the behavior of  $(P_{k}^{} 
- P_{k}^{'})$ for ($1/2,1$) system. It is observed that $(P_2 - P_{2}^{'})$
changes sign within the range $[0,J_c]$, indicating that every condition 
of Eq. (\ref{wcsP}) may
not hold in this range, especially at high bath temperatures. 
However, we observe that the 
majorization conditions continue to hold and $\{P_{k}\} \prec \{P_{k}^{'}\}$,
even if $P_{2}^{'} > P_2$ (see Fig. \ref{figmaj}).
 \subsection{Energy level ordering}
The actual arrangement of the energy eigenvalues  
depends on the positive work conditions 
derived above. As for the relative position of $2sB$ energy level,
it will not change, 
because it is the highest 
energy eigenvalue of the system
regardless of the coupling strength $J$.
The ground state or the minimum
energy state will be decided as follows.

There are two energy levels $-2sB$ and $-2(s-1)B-8sJ$
which can possibly form the ground state of the coupled system,
and their energy gap is $|2B-8sJ|$.
Given that $B_{1}>B_{2}$ and $0<J<J_c$, we can check that
\begin{equation}
  J<J_{c}<
  \frac{B_{2}}{4s}<\frac{B_{1}}{4s}.  
  \label{level}
\end{equation}
The above implies that $2B-8sJ>0$, thereby making $-2sB$
as the lowest energy of the system and $-2(s-1)B-8sJ$ 
as the energy of the first excited state. 
Now, Eq. (\ref{level}) opens different 
possibilities for the arrangement of other energy 
levels. For example, the levels
$-2(s-2)B-8sJ-8(s-1)J$ and $-2(s-1)B$
have an energy gap $|-2B+8sJ+8(s-1)J|$, and
either of them can be at higher
energy state than the other, and both the 
arrangements are acceptable. 
For concreteness, we assume the condition that
there is no level crossing when $B_{1}$ 
is changed to a lower value $B_{2}$.
One way of arranging the 
energy levels, in accordance with Eq. (\ref{level}), 
is shown in Fig. \ref{fig:1}, which is
assumed for the discussion that follows.

The net entropy production in one cycle 
$\Delta S_{av}$ for the coupled system, 
$\Delta S_{av} = -{Q_{1,av}}/{T_{1}}+
{Q_{2,av}}/{T_{2}}$, can be written as:
\be
\Delta S_{av} 
=2X\bigg(\cfrac{B_{2}}{T_{2}}
-\cfrac{B_{1}}{T_{1}}\bigg)+8JY\bigg(\cfrac{1}{T_{2}}
-\cfrac{1}{T_{1}}\bigg).
\ee
In the above expression, due to  Eq. (\ref{pwc}), 
the first term is always positive, 
but since $T_{1} > T_{2}$, the sign of the second term depends on $Y$
which may not be positive.

We will  
consider the WCS whereby
 under Eq. (\ref{WCS1}), 
all terms in the defining sum $Y$ (Eq. (\ref{xy})) are negative,
thus making $Y$ negative definite (note that $m_{2} >0$ for all $k$). 
 Defining
\[Y_{1} =-Y/s, \quad a=2\bigg(\cfrac{B_{2}}{T_{2}}-
\cfrac{B_{1}}{T_{1}}\bigg) >0 
\quad b=8sJ\bigg(\cfrac{1}{T_{2}}-
\cfrac{1}{T_{1}}\bigg) >0, 
\] 
we have,
$ \Delta S_{av} = aX-bY_{1}$.
The condition, given by Eq. (\ref{jc0}), 
on the coupling strength which ensures $W_{av}>0$,
implies that $a >b$. Then, for $Y_1 >0$,   
we have shown in Appendix D that PWC 
for the coupled system encapsulated 
in Eqs (\ref{pwc}) and (\ref{jc0}) 
suffice to prove $X>Y_{1}$ and hence 
$\Delta S_{av}>0$. This establishes 
the consistency of our engine 
with the second law, in the considered domain.

\section{Efficiency enhancement and the upper bound}
In the above, we have established  conditions
for work extraction in the quantum Otto cycle for the coupled system 
 and verified consistency 
with the second law. In this section, we explore 
how the coupling between the spins may enhance the  
efficiency of the engine.

\noindent 
The heat absorbed from the hot reservoir is given by:
$Q_{1,av} = 2B_1 X + 8JY$, where $X$ and $Y$ are as defined in Eq. (\ref{xy}).
From  the energy levels diagram, it is clear that the contribution
$8JY$ to the exchanged heat comes solely from levels which depend on
parameter $J$, apart from the field $B$. 
Now, since, $Q_{2,av} = 2B_2 X + 8JY$, this 'extra' contribution to 
heat is not available for conversion into work, and is wasted
if $8JY >0$.
However, it may be utilized to enhance the 
efficiency of the cycle if $8JY <0$, 
thus effectively decreasing the 
heat absorbed from the hot reservoir. 
Remarkably, the WCS considered earlier implies that
all terms entering the sum for $Y$ are negative, and so
with $J>0$, we have $Y \le 0$. Thus, the WCS directly leads to regime where
we can expect an enhancement of the efficiency.
Thus, for the operational regime discussed in previous
sections, we can rewrite the expression for
efficiency, $\eta = 1-Q_{2,av}/ Q_{1,av}$ as follows.
\begin{equation}
   \eta =\cfrac{\eta _{0} }
{1 + \cfrac{8J Y}{2XB_{1} } } =
\cfrac{\eta _{0} }{1-\cfrac{4sJ Y_{1}}{XB_{1} } } 
\label{eta1} 
\end{equation}
where $Y_{1} =-Y/s >0$. 
We have proved in Appendix D that 
$X>Y_{1}$. With
$B_{1}>4sJ$ (Eq. (\ref{level})), we  obtain 
\be
\eta < \cfrac{\eta _{0}}
{1-4sJ/B_{1} } <1-\cfrac{T_{2} }{T_{1} } =\eta _{C},
\ee
where the second inequality follows due to the permissible
range of $J$ (Eq. (\ref{jc0})).
Thus, the expression
\begin{equation}
\eta_{ub} = \cfrac{\eta _{0} }{1- {4sJ}/{B_{1}} }
\label{etaub}
\end{equation}
constitutes an upper bound to the system's 
efficiency which is tighter than Carnot efficiency, and 
within the coupling range $0<J<J_{c}$.

The above expression bounding the efficiency of Otto cycle
is our main result of the paper. This expression is validated
with numerical calculations in the discussion section.
Note that $\eta_{ub}$ given by Eq. (\ref{etaub}) 
is dependent solely on the field values and the total spin 
of the two particles while it is independent of the bath 
temperatures. This expression generalizes the 
upper bound derived earlier in Ref. \cite{thomas2011coupled}
for the $(\frac{1}{2},\frac{1}{2})$ system.

We close this section with a remark on 
the three possible spin combinations for our ($s_1, s_2$) system.   
\begin{itemize}
	\item when one spin value
	is half-integral and other is integral
	\item when both valus are 
	half-integral or both are integral
	\item when both are 
	of the same magnitude 
	(both as  half-integral or integral)
\end{itemize}
\noindent In this paper, we have discussed the first case only.
The only difference
between the present case and the other two cases
is that for the latter, when the spins are uncoupled, 
an energy level with zero energy and $2s_{1}+1$-fold
degeneracy occurs but that does not affect
the performance of the system. The reason
is that after the coupling is turned on between 
the spins, this energy state splits into 
$2s_{1}+1$ non-degenerate energy levels
which depend only on the coupling factor $J$. 
Since $J$ is kept fixed during the cycle, therefore 
these levels do not shift in a 
cycle and hence do not contribute to the
average work resulting in the same PWC 
as already derived for the first case. 
Similarly, it can be seen that these levels
do not change the condition for maximal
efficiency enhancement and same upper bound 
can be obtained, whatever be the spin combination.

\section{Complete Otto Cycles}
The working medium for the classical Otto cycle is usually a
macroscopic system amenable to thermodynamic
treatment. This medium may be
a collection of statistically independent, non-interacting
individual quantum systems or {\it elements}, such as spin-$1/2$ particles,
or harmonic oscillators and so on. In the adiabatic step
of the Otto cycle, the thermodynamic entropy of the
working medium stays constant.
This implies that there
is no intrinsic control on the transitions experienced by
individual elements of the working medium.

On the other hand, the working medium of a quantum Otto engine
consists of individual elements.
In a quasi-static cycle,
the isochoric steps are
stochastic while the adiabatic steps are deterministic.
The quantum adiabatic
step is executed slowly enough such that no
transition is induced between energy levels of the element
which continues to occupy its initial state throughout the process.
Thus at the level of the ensemble, the occupation
probabilities do not change during this process.
Such a process thus imposes maximal control on the
evolution of the isolated element, and it is described
by a quantum unitary process.

Still, due to the stochastic nature of the
contact with the reservoirs, the element may not return to its
initial state, after the four steps of the cycle.
Usually, we are interested in the
average properties of the cycle by which the
quantities like heat and work are defined at
the ensemble level. In this section, we focus
on the complete Otto cycles (COCs) inherent in the 
average Otto cycle considered in earlier sections. 
The reason that  Otto cycle is so often studied
in the quantum thermodynamics literature is that
the contributions towards heat and work can be
clearly separated into different steps---which helps in
the analysis. This distinction also holds at the level
of COCs; the interaction of the working medium with a 
reservoir involves
only exchange of heat with the reservoir, whereas the 
quantum adiabatic step involves only work.

\begin{figure}      
\begin{center}
\includegraphics[width=
8cm,height=8cm,keepaspectratio]{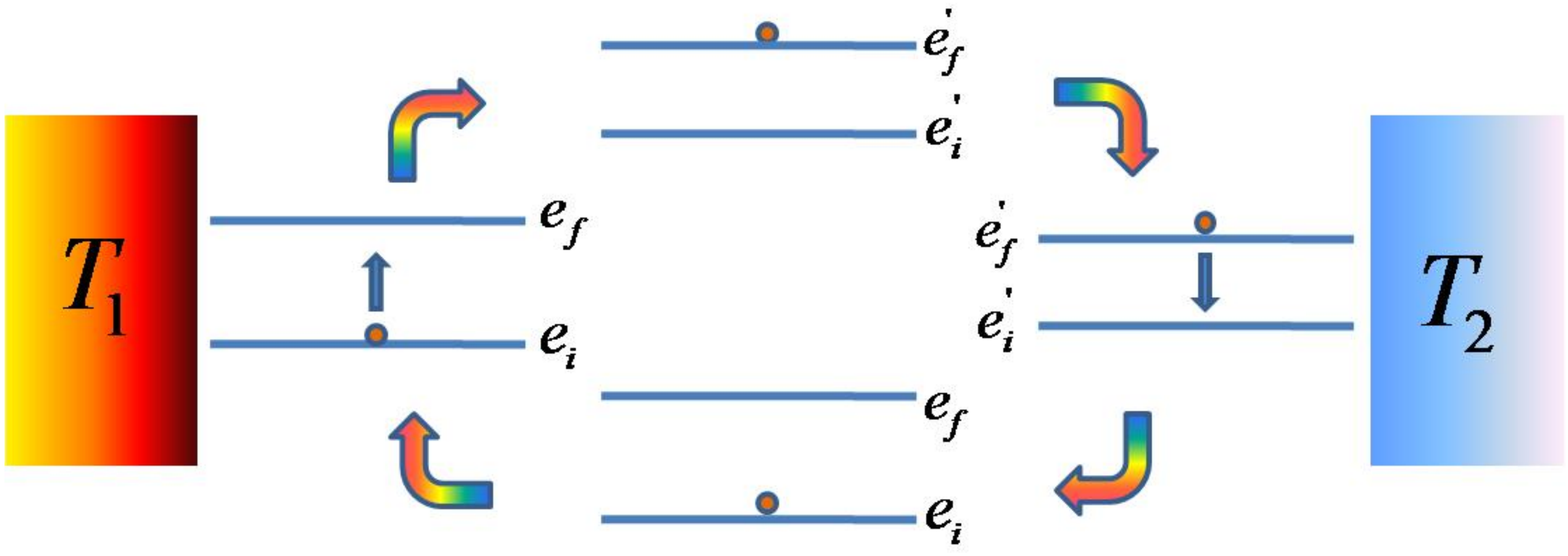}\\
\caption{Schematic of a 
complete Otto cycle (COC) as an engine using two heat 
reservoirs ($T_1 > T_2$), involving two energy levels 
 of the working medium. The
heat absorbed from the hot reservoir is $q_1 = e_{f}-e_{i}$,
while the heat rejected to the cold bath is 
$q_{2} = e_{f}^{'}-e_{i}^{'}$. The work 
extracted per complete cycle is $w = q_1 - q_2$.
}
\label{fig:4}      
\end{center}
\end{figure}

Consider, the COC shown as an engine in Fig. \ref{fig:4}. 
If the working medium starts in energy level $e_{i}$,
then by the end of the four stages, it
is again found in level $e_{i}$. Such a cycle can
either run forward as an engine,  
or backwards as a refrigerator.
Analysing the 
performance of COCs is much easier since
we are dealing with only 
two levels at a time without  
invoking occupation probabilities
of the levels and any average quantities.
 Let us represent an energy eigenvalue of
 the uncoupled system as, $e \equiv m_{1}B$, 
 where $m_{1}$ varies from $ m_{1}=-2s,...,+2s$.
 Based on the final ($f$) and initial ($i$) values of $m_1$, 
 let us define the quantity $x = m_{1,f}-m_{1,i}$, ranging as 
 $x = \pm 2, ...,\pm 4s$.
 Let $q_{1},q_{2},w$ respectively denote the heat exchanged with the 
 hot bath, cold bath and the work performed:  
 \bea
  q_{1} &=& e_{f}-e_{i}=xB_{1}, \nonumber \\
  q_{2} &=&e_{f}^{'}-e_{i}^{'} =xB_{2}, \\
  w &=& q_{1}-q_{2}= x(B_{1}-B_{2}). \nonumber
 \eea
 With $B_{1}>B_{2}>0$, we have $q_h, q_c >0$ and $w>0$, if $x>0$.
It is clear that for $x>0$ ($x<0$), a COC runs as an engine (refrigerator).
 The net entropy change ($\Delta S_0$) is 
 contributed only by the reservoirs.
 Thereby, we obtain
 \begin{equation}
\Delta S_{0}=-\frac{q_{1}}{T_{1}}+
\frac{q_{2}}{T_{2}}=x\left(-\frac{B_{1}}{T_{1}}
+\frac{B_{2}}{T_{2}}\right).
\label{s}
 \end{equation} 
Now, for $x>0$,  
the condition ${B_{2}}/{T_{2}} > {B_{1}}/{T_{1}}$
ensures that $\Delta S_{0}>0$, or we may say that 
the second law
is then satisfied at the level of COC. 
Note that there is a subtle difference in the statement 
about the second law at the level of a COC versus the average performance level.
In the former case, $x>0$ guarantees the operation of an
engine, whereas the {\it additional} condition,
${B_{2}}/{T_{2}} > {B_{1}}/{T_{1}}$, 
makes this operation consistent with the second law.
On the other hand, for the average operation as an engine,
we require $v>0$
which itself requires the condition (\ref{pwc}).
The latter then automatically
ensures consistency with the second law 
at the level of average performance.

Also note that
we do not impose the second
law at the level of a COC, and
the net entropy change for a COC may be negative,
as for instance, with $x<0$ or a COC operating
as a refrigerator, if ${B_{2}}/{T_{2}} > {B_{1}}/{T_{1}}$.
Thus, we do not imply that COCs with $\Delta S_0 <0$
do not happen. These observations
lead to the following interesting conclusion about
the uncoupled model.
{\it A consistency with the second
law for the average performance as engine 
ensures consistency with the second law
for a COC as engine, and vice versa}. 

Let us study the effect of coupling between the spins.
  Now, there are no degenerate levels.
  Expressing an energy eigenvalue of the coupled system as,
 $E \equiv m_{1} B-8m_{2} J$, where
 $ m_{1},m_{2}$ values are given in Table \ref{table:5}.
 The levels with same $m_1$ were
 originally degenerate in the uncoupled model.
 For the coupled model, 
 energy levels belong to the
 same band if they have the 
 same value of $m_{1}$, but have different
 values of $m_{2}$. 
 Also note that in every band, there is one
 level that stays at 
 the same energy even after the coupling is switched on.
 
 Now, for a COC between any two energy
 levels of the coupled system, the general forms of 
 heat exchanged with the reservoirs,
  $Q_{1}$, $Q_{2}$, and the work performed, $W= Q_1 - Q_2$,
  can be written as
\bea
  Q_{1} &=& xB_{1}+8Jy, \nonumber \\
Q_{2}  &= & xB_{2}+8Jy, \\
   W  &=& x(B_{1}-B_{2}), \nonumber
\eea   
with  $x = m_{1,f}-m_{1,i}$ and   $y = m_{2,f}-m_{2,i}$.
The net entropy change in one cycle is
\begin{equation}
\Delta S =-\frac{Q_{1}}
{T_{1}}+\frac{Q_{2}}{T_{2}}=
x\left(\frac{B_{2}}{T_{2}}-
\frac{B_{1}}{T_{1}}\right)+8Jy 
\left(\frac{1}{T_{2}}-\frac{1}
{T_{1}}\right).
\label{s1}
\end{equation} 
We discuss the possible COCs as below.
\\1. ${\bf x \ne 0,y=0}$:  
These cycles occur between
any two different energy bands having the same $m_{2}$.
Therefore, if such a cycle proceeds 
as an engine ($x>0$), its efficiency is 
$W/Q_{1}=1-B_{2}/B_{1} = \eta_{0}$. 
From Eq. (\ref{s1}), this COC is
consistent with the second law, for $B_{2}>B_{1}\theta $.
\\2. ${\bf x=0, y \ne 0}$: 
These cycles are possible between energy levels
of the same band i.e having same $m_{1}$.
The work performed is zero, and
the heat exchanged is, 
$Q_{1}=8J y = Q_{2}$. Thus, for $y>0$, 
the corresponding
efficiency is also zero. 
 \\3. ${\bf x,y \ne 0}$ {\bf with the same sign}: These cycles
are possible between different bands for levels with 
different $m_{1}$ and $m_{2}$. 
If such cycles proceed as engine
i.e $x>0$ (and $y>0$),
then the corresponding efficiency is 
\be
\eta_{}=\cfrac{\eta_{0}}{1+\cfrac{8yJ}
{xB_{1}}}<\eta_{0}.
\ee
From Eq. (\ref{s1}),
this type of COC is consistent with the second 
law for $B_{2}>B_{1}\theta$,  {\it without}
imposing any further
condition on the coupling strength $J \geq 0$.
Therefore, if the second law allows COCs with
$\eta=\eta_{0}$, then it also allows COCs with $\eta<
\eta_{0}$.  \\
4. ${\bf x,y \ne 0}$ {\bf with opposite signs}:
 These cycles occur between 
energy levels of different bands with 
different $m_{1}$ and $m_{2}$. 
If $x>0$ for such cycles (and  $y<0$), 
the corresponding efficiency is
\begin{equation}
\eta=
\cfrac{\eta_{0}}{1-\cfrac{8|y|J}{xB_{1}}}>\eta_{0}.
\label{et}
\end{equation} 
From Eq. (\ref{s1}), 
this COC is allowed 
by the second law, if $B_{2}>B_{1}\theta $ and 
\begin{equation}
0 < J < \cfrac{x(B_{2}-B_{1}\theta)}
{8|y|(1-\theta)} \equiv J_{a}.
\label{ja}
\end{equation}
Now, we look for the values of $x$ and $y$ which place the most
stringent condition on the second law (Eq. (\ref{s1})), 
or,  in other
words, which make $\Delta S$ as the least positive.
This will be the worst-case scenario (WCS) in this context, as 
other values of $x$ and $y$ would yield a larger
upper bound $J_a$. Thus, the range 
imposed by the WCS will hold for {\it all} COCs,
making all of them consistent with the second law.

The first term in Eq. (\ref{s1}) takes the minimum
value if $x=2$.
For the second term, let $y_{\rm min} <0$ denote 
the minimum value of $y$. Then, we obtain
$-y_{\rm min}=[s+(s-1)+..
+(s-(2s_{1}-1))]=s_{1}(2s_{2}+1)$. 
Substituting the above values of $x$ and $y$ in
Eq. (\ref{ja}), we obtain 
the following range of $J$:
\be
0<J< \frac{B_{2}-B_{1}\theta}{4s_{1}(2s_{2}+1)(1-\theta)} \equiv J_{x}.
\ee 
Therefore, it follows that for $B_{2}>B_{1}\theta$
and within the range $0<J<J_{x}$, \textit{ all} 
the COCs perform as an engine and satisfy the second law.

Now, from the probabilistic or average analysis, 
we concluded that the conditions $B_{2}>B_{1}\theta$
 and the coupling range $0<J< J_{c}$, 
ensure the average
performance as an engine.
To compare the two ranges for $J$,
note that $s_1(2 s_2 +1) \geq s$, 
where the equality is obtained for $s_1 = 1/2$ 
implying that, in general, $J_{x} \leq J_{c}$.
This has the following important consequence.
The range for the parameter $J$, in which the 
machine behaves as an engine on average,
subsumes the range for $J$ in which 
all COCs, performing as an engine, are also 
consistent with the second law. 
Conversely, if we restrict to the range $0< J < J_x$,
allowing all COCs running as engine to
follow the second law, then   
the average operation as an engine, in that range
of parameters, is also consistent with the second law.

Also, from Section V, we learn that out of all the
possible  COCs with $\eta>\eta_{0}$,
the maximum possible value of 
efficiency is obtained  
from  Eq. (\ref{et}) for minimum $x$ 
i.e $x=2$ and $\vert y_{min}\vert= s_{1}(2s_{2}+1)$, given by
\begin{equation}
\eta_{\rm max}=\cfrac{\eta_{0}}
{1-\cfrac{4s_{1}(2s_{2}+1)J}{B_{1}}}.
\label{emax}
\end{equation} 
This cycle is allowed by the
second law for the condition
$B_{2}>B_{1}\theta$ and in the $0<J<J_{x}$ 
range of coupling. 
 Interestingly, the coupling range required for $W_{av}>0$ goes
beyond $J=J_{x}$, since 
$J_{x} \le J_{c}$. The case of $J_a=J_{c}$ 
is obtained when we substitute 
$x=2,|y|=s$ in Eq. (\ref{ja}),
and out of all the COCs allowed
in this range, the maximum efficiency
is given as, 
${\eta_{0}}/
({1-{4sJ}/{B_{1}}})$. The latter value is 
same as the upper bound,
$\eta_{ub}$, inferred 
by analysing the average performance 
of the system. As can be seen,  
$\eta_{ub} \leq \eta_{max}$.
For the special case of $(1/2,s_2)$ working medium,  
 $J_{x}$ and $J_{c}$ values 
coincide irrespective of the value of $s_{2}$, 
leading to $\eta_{max}=\eta_{ub}$.
\section{Discussion}
We have analyzed the performance of a quantum 
Otto engine based on a working medium with a complex
energy spectrum. An insight into the possible operational
regimes is hard to obtain analytically for such a system.
Using a heuristic-based approach 
and employing techniques such as
worst-case/best-case reasoning, we have highlighted a regime 
in which the machine definitely works as an engine on average. 
These set of conditions can be related to 
the concept of majorization for the given model. 
Thereby, we find that majorization serves as a more robust
criterion for positive work extraction from our engine.

We also introduced an analysis based on  complete Otto
cycles (COCs). Compared to the probabilistic
analysis, the COC approach is much simpler and straightforward.
The latter utilizes
much less information than the 'average' analysis,
and the conclusions so obtained may not be as general.
However, as a starting point,
the criteria for COCs may serve  
as a useful heuristic to gain insight into the average 
performance of the  Otto machine. As we have seen,
there is an interesting correspondence between 
the COCs and the average Otto cycle with regard to the 
validity of the second law.
\begin{figure}      
\begin{center}
\includegraphics[width=
7cm,height=7cm,keepaspectratio]{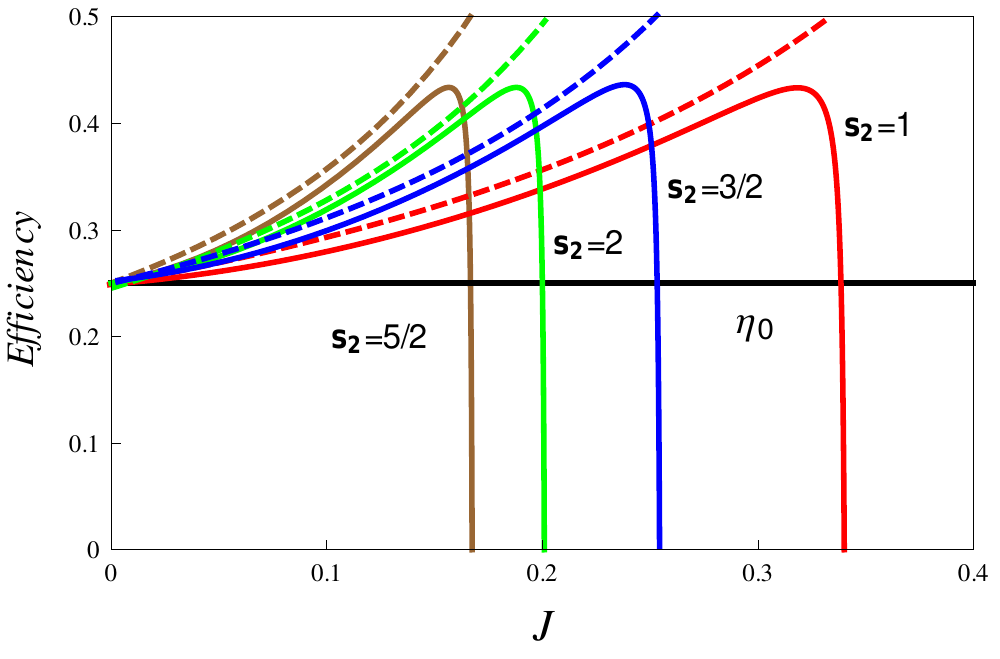}
\caption{Variation of efficiency (solid lines) for different 
values of spin $s_2$, 
with $s_{1}=1/2, B_{1}=4, B_{2}=3, T_{1}=1$ and $T_{2}=0.5$. 
The corresponding upper bounds ($\eta_{ub}$) have been shown 
by dashed lines. 
The uncoupled efficiency 
($\eta_{0}$) shown by horizontal black line.}
\label{fig:5}      
\end{center}
\end{figure}
 One of our main results is an explicit expression for the upper bound 
 of Otto efficiency for the coupled system. This expression reduces
 to the one found for ($1/2,1/2$) case, with $s=1$ 
 \cite{thomas2011coupled},
 or to the case of coupled, effective two-level systems 
 \cite{mehta2017quantum}.
 The dependence of the average efficiency on coupling factor $J$  
 and validity of the upper bound is demonstrated in Fig. \ref{fig:5}.
 
 \begin{figure}
\subfigure[]{\includegraphics[width=
7cm,height=8cm,keepaspectratio]{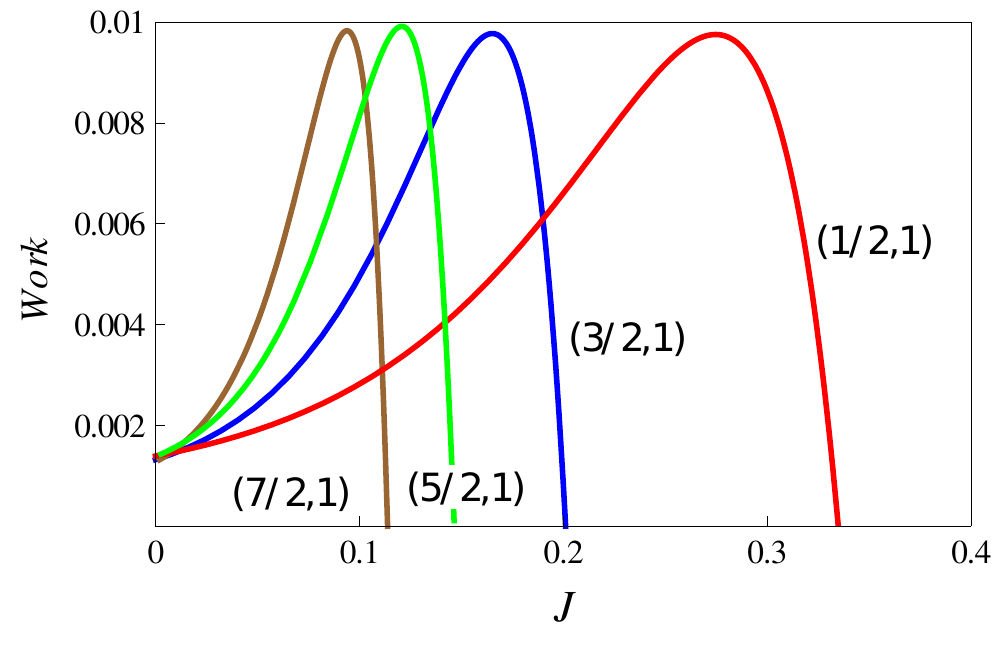}} 
   \subfigure[]{\includegraphics[width=
7cm,height=7cm,keepaspectratio]{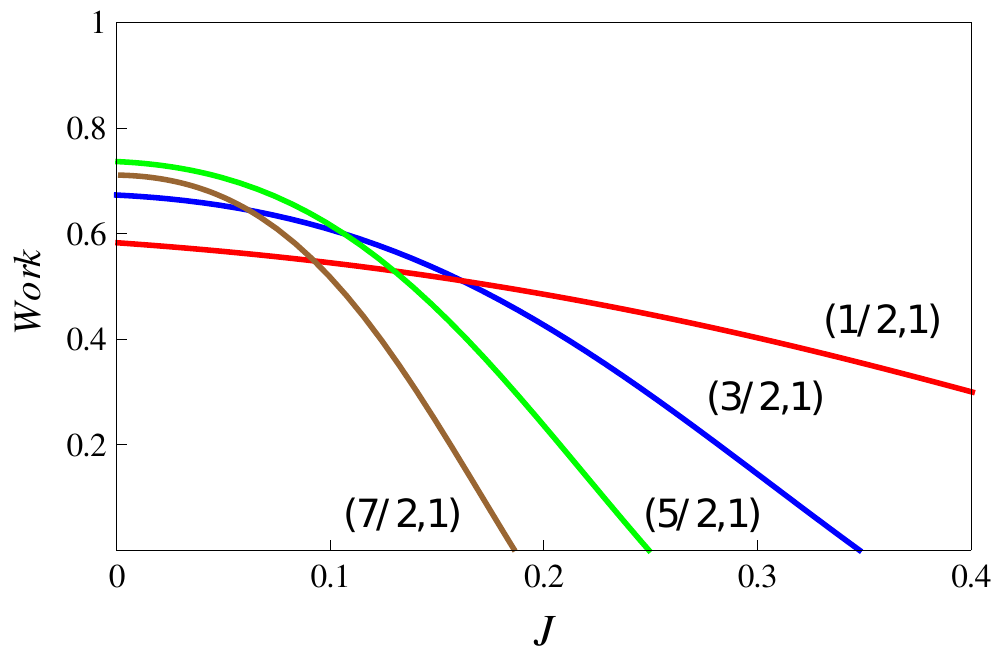}}
    \caption{Variation of extracted work with coupling strength $J$ for 
    different spin combinations ($s_{1},s_{2}$). The fields are set at 
    values $B_1=4, B_2=3$ and the bath temperatures are: (a) 
$T_{1}=1$,$T_{2}=0.5$ (b) $T_{1}=6,T_{2}=3$.}
    \label{fig:6}
\end{figure}
\begin{figure}
  \subfigure[]{\includegraphics[width=
7cm,height=7cm,keepaspectratio]{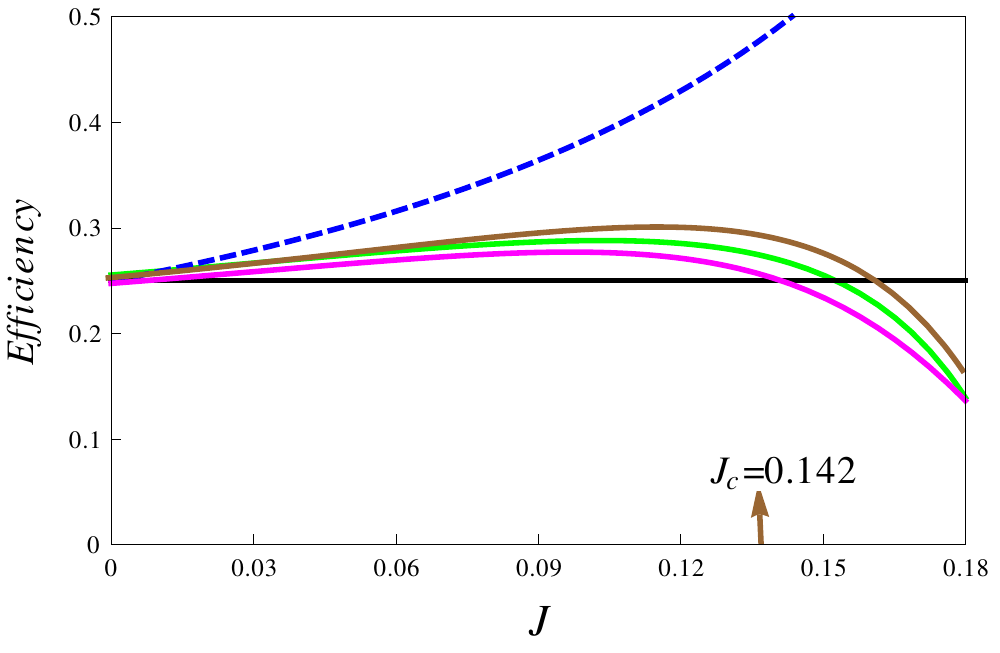}} 
    \subfigure[]{\includegraphics[width=
7cm,height=7cm,keepaspectratio]{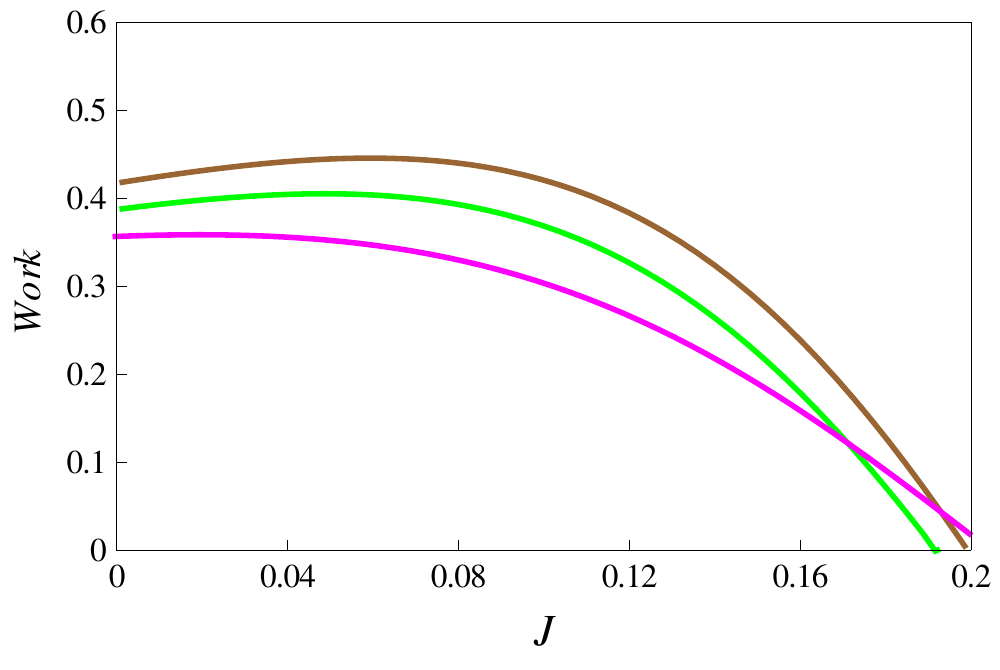}}
\caption{Variation of (a) efficiency and (b) work with coupling strength 
$J$ for different spin combinations ($s_{1},s_{2}$) where $s=7/2$ is held fixed. 
The solid pink, green and brown lines respectively show the variation for 
$(1/2,3), (1,5/2)$ and $(3/2,2)$ cases. The parameters are set at values 
$B_1=4, B_2=3$,   $T_{1}=4,T_{2}=2$. Here, $J_{c}=0.142$. The upper bound and 
the uncoupled efficiency are respectively shown by dashed blue and black lines 
in (a). The Carnot efficiency is $0.5$.}
    \label{fig:7}
\end{figure}

Besides the above analytic approaches,   
 we may also numerically study the implications of using
 higher spins on the performance of thermal machines.
To make a few observations, we note that the higher $"s"$ values shift the 
maximum of work to the weak coupling regimes as shown in 
Fig. \ref{fig:6}a. Thus, higher magnitudes of spin may be a useful resource
to achieve more work output for weak coupling strengths.
Numerical analysis also 
shows that increasing the bath temperatures may increase
the work output by orders of magnitude (see Fig. \ref{fig:6}b). We also 
observe an extended regime of positive work extraction from the
system at high temperatures and this effect is more pronounced for 
lower "$s$" values. 
Along these lines, 
 variations of the efficiency and work output with the coupling factor $J$,
 may be studied where $s_1$ and $s_2$ are
 varied for a fixed $s$ value. Fig. \ref{fig:7} shows
 different cases for the case of $s=7/2$. Note that $\eta_{ub}$ and 
 $J_{c}$ (which depend on $s$ and not on the values of individual spins) are
 same for a given $s$. 
\par
 Finally, other possible domains of operation such as refrigerator and 
 accelerator may be addressed using the techniques explored in this paper.
 The study of local thermodynamics of 
 individual spins relative to the global performance, and  
 other models of coupled spins featuring different  
interactions are some of the potential avenues of future inquiry.

   %
\appendix
\renewcommand{\thesubsection}{\Alph{section}.\arabic{section}}
\appendix
\label{Appendix}
\numberwithin{equation}{section}
\section{PWC for the uncoupled model}
\noindent The net work extracted from the system when 
$s_{1}$ and $s_{2}$ are uncoupled, is given as,
$w_{av}=2(B_{1} -B_{2} )v$. Since we assume $B_{1} >B_{2}$, 
we need to find conditions for 
$v>0$ to hold, i.e. 
\begin{equation}
v = \left\{\begin{array}{l}
s\left(p_{1}^{'}-p_{1} +p_{n}-p_{n}^{'} \right)
+\\ (s-1)\left(\begin{array}{l} p_{2}^{'} -p_{2} +p_{3}^{'}
-p_{3}+p_{n-2}-p_{n-2}^{'}+p_{n-1}-p_{n-1}^{'}
\end{array}\right)+...+\\ \left(s-r\right)
\left(\begin{array}{l} p_{n/2-2s_{1}}^{'} 
-p_{n/2-2s_{1}}...+p_{n/2}^{'}-p_{n/2}+
\\ p_{n/2+1}-p_{n/2+1}^{'}+...+p_{n/2+2s_{1} +1}
-p_{n/2+2s_{1} +1}^{'} \end{array}\right)
\end{array} \right. >0,
\label{proof}
\end{equation}
where $r=s-1/2$. Let us denote the
term in $v$ with the largest coefficient $s$,  as
$L \equiv (p_{1}^{'}-p_{1}+p_{n}-p_{n}^{'})$.
We will show that  $L<0$ implies $v<0$ or
in other words if the term with largest coefficient 
in $w_{av}$ is negative, 
the system cannot work as an engine.
\begin{center}
\textbf{Lemma 1}: $\mathbf{ L<0}$
\textbf{implies} $\mathbf{v<0}$ 
\end{center}
Let us look at the explicit expression of 
$L \equiv (p_{1}^{'}-p_{n}^{'})- (p_{1}-p_{n})$:
\be
{L}=\cfrac{e^{2sB_{2}/T_{2}}
-e^{-2sB_{2}/T_{2}}}{z_{2}(B_{2}/T_{2})}-
\cfrac{e^{2sB_{1}/T_{1}}-e^{-2sB_{1}/T_{1}}}
{z_{1}(B_{1}/T_{1})},\ee
where $z_i(B_i/T_i)$ is the
partition function for the system given by Eq.
(\ref{z1}). The above expression is of the form, 
\begin{equation}
{L}= f(B_{2}/T_{2})-f(B_{1}/T_{1}).
\label{fbt}
\end{equation} 
Let us observe the function 
$f(B_{1}/T_{1})=p_1-p_n$.
First, due to canonical form of probabilities,
we know that $p_1 > p_n$, and so $f(B_{1}/T_{1}) >0$.
Then, for a given value $B_{1}$, if we increase the temperature 
$T_{1}$, thereby decreasing 
$B_{1}/T_{1}$, we know that the difference
$p_{1}-p_{n}$ decreases and vice versa.
This implies that $f(B_{1}/T_{1})$
is a monotonically increasing
function of $B_{1}/T_{1}$. The
same is also true for $f(B_{2}/T_{2})$.\\
Since $f({B/T})>0$ is a monotonic
increasing function of $B/T>0$, 
so if $L<0$, the following condition must hold:
\begin{equation}
\cfrac{B_{2}}{T_{2}}<\cfrac{B_{1}}{T_{1}}.
\label{lemma1}
\end{equation} 
The above condition further implies $z_{2}<z_{1}$, and so
\begin{equation}
p_{n}^{}-p_{n}^{'}= 
\cfrac{e^{-2sB_{1}/T_{1}}}{z_{1}}-
\cfrac{e^{-2sB_{2}/T_{2}}}{z_{2}}<0,
\label{z2z1}
\end{equation}
\begin{equation}
p_{1}^{'}-p_{1} =
\frac{1}{\sum_{l=0}^{2s_{}}
g_{|s-l|} e^{-2lB_{2} /T_{2} }  }
-\frac{1}{\sum_{l=0}^{2s_{}} 
g_{|s-l|}e^{-2lB_{1} /T_{1} }}<0.
\end{equation} 
We now rewrite the expression of $v$ as follows
\begin{equation}
v= \left\{\begin{array}{l} 
s\begin{array}{l} 
\left[p_{1}^{'} \left(1-e^{-4sB_{2} /T_{2}}
\right)-p_{1}\left(1-e^{-4sB_{1}/T_{1} }\right)\right]
\end{array}+\\ (s-1)\begin{array}{l}
\left\{\left(p_{2}^{'} +p_{3}^{'} \right)
\left(1-e^{-4(s-1)B_{2} /T_{2} } \right)-\left(p_{2}+p_{3}\right)
 \left(1-e^{-4(s-1)B_{1} /T_{1} } \right)\right]\end{array}+
 ...+\\ \left(s-r\right)\left[\left(p_{n/2-2s_{1}}^{'}
 +...+p_{n/2}^{'}\right)\left(1-e^{-2B_{2} /T_{2}} \right)
 -\left(p_{n/2-2s_{1}}+...+p_{n/2}\right)
 \left(1-e^{-2B_{1} /T_{1} }\right) \right].  
 \end{array} \right.
\label{v2}
\end{equation}
\noindent 
Now, if $L<0$ and so Eq. (\ref{lemma1}) holds, then
in the first term above, accompanying the coefficient $s$, 
we have  
\[1-e^{-4sB_{2} /T_{2} }<
1-e^{-4sB_{1} /T_{1} }.\]
Similarly, in the second term of the expression for $v$,
\[1-e^{-4(s-1)B_{2} /T_{2} }
<1-e^{-4(s-1)B_{1} /T_{1} },\]
and so on, till we have
\[1-e^{-2B_{2} /T_{2} } 
<1-e^{-2B_{1} /T_{1} }, \]
in the last term. 

It is important to note that
Eq. (\ref{lemma1}) does not
imply any definite relation between 
$p_{k}$ and $p_{k}^{'}$ 
for $k=2,...,n/2$. On the other hand, it is clear
from Eq. (\ref{v2}) that $p_{k}^{'}>p_{k}$ for all 
$k=2,..,n/2$, would favor the case $v>0$.
So, assuming $L<0$, we will now consider
the BCS (Best Case Scenario), mathematically written as, 
\be
p_{k}^{'}>p_{k} \quad \forall \quad k=2,..,n/2, 
\ee
and show that $v<0$. The proof is as follows.\\

{\bf Proof:} It has been noted earlier that $L<0$ implies Eq. 
(\ref{lemma1}) and $z_{2}<z_{1}$. 
From inspection of the form of canonical probabilities, 
this further leads to $p_{k}<p_{k}^{'}$,
 $\forall$ $k=n/2+1,...,n$. 
Using these relations 
in the normalization condition of probabilities given as, 
$\sum_{k=1}^{n}(p_{k}^{'}-p_{k})=0,$  
we have, $s\sum_{k=1}^{n/2}(p_{k}^{'}-p_{k})<0$ 
along with the following:
\[(s-r)(p_{n/2+1}-p_{n/2+1}^{'})<0, \quad ... \quad 
(s-1)(p_{n-1}-p_{n-1}^{'})<0, \quad s(p_{n}-p_{n}^{'})<0. 
\] 
Also, under BCS, we have
\[(-1).\left(p_{2}^{'}-p_{2} \right)<0, \quad (-1).
\left(p_{3}^{'}-p_{3} \right)<0, \quad ..., \quad (-r).
\left(p_{n/2}^{'}-p_{n/2} \right)<0.\] 
Adding all the above inequalities, we arrive at the 
result $v<0$, thereby proving \textit{Lemma} 1.

\begin{center}
\textbf{Lemma 2}: $\mathbf{ L>0}$
\textbf{implies} $\mathbf{v>0}$   
\end{center} 
Using the monotonic property of $L$,
it is obvious that if $L>0$, the following must hold:
\begin{equation}
\cfrac{B_{2}}{T_{2}}>\cfrac{B_{1}}{T_{1}}.
\label{cond}
\end{equation} 
It can be seen that the above condition
implies $p_{1}^{'}>p_{1}$ and $p_{n}>p_{n}^{'}$.
Eq. (\ref{cond}) favors $v>0$ as it leads
to the condition 
\[
1-e^{-2 m_1  B_{2} /T_{2} }
>1-e^{-2 m_1 B_{1} /T_{1} },
\] 
in all the  
terms in Eq. (\ref{v2}), as 
$m_1 >0$ for all the upper-half levels i.e for $k=n/2+1,...,n$ (see Table 3).
Also under Eq. (\ref{cond}), positivity of Eq. (\ref{v2})
is always favored 
irrespective of the relation between 
$p_{k}$ and $p_{k}^{'}$ for all $k=n/2+1,...,n$. 
As for the rest of the 
occupation probabilities, Eq. (\ref{cond}) 
does not imply any relation between them except
for $p_{1}^{'}>p_{1}$. But it is obvious from
Eq. (\ref{v2}) that $p_{k}^{'}<p_{k}$ for 
all $k=2,..,n/2$ would {\it not} favor $v>0$.

So, assuming $L>0$, we will now consider 
the WCS (Worst Case Scenario), 
mathematically written as,
\be
p_{k}^{'} < p_{k} \quad \forall \quad k=2,3...,n/2,
\ee 
and then show $v>0$. This would prove \textit{Lemma 2}.

{\bf Proof:} 
The condition $L>0$ yields
Eq. (\ref{cond}), leading to $z_{2}>z_{1}$ 
which further implies 
\begin{equation}
  p_{k}^{'} < p_{k}, \quad k=n/2+1,...,n.
\label{pk}
\end{equation}
Thus, we can write
\begin{equation}
  p_{k}^{'} < p_{k}, \quad k=2,...,n.
\label{pk1}
\end{equation} 
These inequalities, along with the 
normalization of each probability distribution, imply 
\begin{equation}
  p_{1}<p_{1}^{'}.
  \label{p1}
\end{equation} 
Now, using Eq. (\ref{pk}) and the 
normalization of probability distributions,
we can write
\[s\sum_{k=1}^{n/2}(p_{k}^{'}-p_{k})>0,
\] 
along with the following conditions:
\[(s-r)(p_{n/2+1}-
p_{n/2+1}^{'})>0, \quad ..., \quad(s-1)(p_{n-1}-p_{n-1}^{'})
>0, \quad s(p_{n}-p_{n}^{'})>0.
\]
Also, under WCS, we have:
\[(-1).\left(p_{2}^{'}-p_{2} \right)>0, \quad 
(-1).\left(p_{3}^{'}-p_{3} \right)>0, \quad ..., \quad
(-r). \left(p_{n/2}^{'}-p_{n/2} \right)>0.
\] 
Adding all the above inequalities, we obtain $v>0$,
thereby proving \textit{Lemma} 2 and concluding that $L>0$, or
Eq. (\ref{cond}), is a 
necessary and sufficient condition for 
positive work extraction from the uncoupled spin system.

\section{PWC for the coupled model}
When the spins are interacting, the work extracted is given as
\[W_{av}=2(B_{1}-B_{2})X,
\]
where 
$X= \cfrac{1}{2}\sum_{k=1}^{n} m_{1}(P_{k}^{}-P_{k}^{'})$. 
The explicit expressions of occupation probabilities
are given in Table \ref{table:2}. With $B_{1}>B_{2}$, 
we need to find the condition for which we have $X>0$,
where
\begin{equation}
X = \left\{\begin{array}{l} s\left(P_{1}^{'}-P_{1} 
+P_{n}-P_{n}^{'} \right)+\\ (s-1)\left(\begin{array}{l}
P_{2}^{'} -P_{2} +P_{3}^{'} -P_{3}+P_{n-2}-P_{n-2}^{'}
+P_{n-1}-P_{n-1}^{'}\end{array}\right)+...+\\
\left(s-r\right)\left(\begin{array}{l}
P_{n/2-2s_{1}}^{'} -P_{n/2-2s_{1}}...
+P_{n/2}^{'}-P_{n/2}+\\ P_{n/2+1}-P_{n/2+1}^{'}+.
..+P_{n/2+2s_{1} +1}-P_{n/2+2s_{1} +1}^{'} 
\end{array}\right).\end{array}\right.
\label{X}
\end{equation}
 \noindent As shown in Appendix A, for the 
uncoupled spins case, the term with the 
largest coefficient ($s$) must be positive,
i.e $L > 0$, for the system to run as
an engine and that is possible if
the system's parameters satisfy Eq. (\ref{cond}).
Now, we are interested to seek additional
conditions which ensure positive work extraction
for the coupled case, provided that the uncoupled model
works as an engine. \\

For completeness, we first show that the same
conditions as (\ref{cond}) also serve as PWC
for the coupled model.
To prove it, consider the term 
$L_{X} \equiv (P_{1}^{'}-P_{1}+P_{n}-P_{n}^{'})$.
We will first show that the
opposite condition, given by Eq. (\ref{lemma1})
yields $L_{X}<0$ and so $X<0$.
Assuming Eq. (\ref{lemma1}), we have 
$Z_{2}<Z_{1}$ as well as
\begin{equation}
P_{n}^{}-P_{n}^{'}=
\cfrac{e^{-2sB_{1}/T_{1}}}{Z_{1}}-
\cfrac{e^{-2sB_{2}/T_{2}}}{Z_{2}}<0.
\label{Z2Z1}
\end{equation} 
Now, it can be seen from
the explicit expressions of
$P_{1}$ and $P_{1}^{'}$ that
for $T_{1}>T_{2}$, Eq. (\ref{lemma1}) 
implies $P_{1}^{'}<P_{1}$.

Therefore we conclude that, 
under Eq. (\ref{lemma1}),  $L_{X}$ is negative definite.
Consider now the expression 
of $X$, rewritten as
\begin{equation}
X = \left\{\begin{array}{l}  s\begin{array}{l}
\left[P_{1}^{'} \left(1-e^{-4sB_{2} /T_{2}}\right)
-P_{1}\left(1-e^{-4sB_{1}/T_{1} }\right)\right]
\end{array}+\\ (s-1)\begin{array}{l} 
\left[\left(P_{2}^{'} +P_{3}^{'} \right) 
\left(1-e^{-4(s-1)B_{2} /T_{2} } \right)-
\left(P_{2}+P_{3}\right)
 \left(1-e^{-4(s-1)B_{1} /T_{1} } \right)
 \right]\end{array}+...+\\ \left(s-r\right)
 \left[\left(P_{n/2-2s_{1}}^{'}+...+P_{n/2}^{'}
 \right)\left(1-e^{-2B_{2} /T_{2}} \right)-\left
 (P_{n/2-2s_{1}} +...+P_{n/2}\right)
 \left(1-e^{-2B_{1} /T_{1} }\right)\right].
 \end{array}\right.
\label{X1}
\end{equation} 
 As can be seen, Eq. (\ref{lemma1}) 
 or $L_{X}<0$ implies, that the following conditions
\[
1-e^{-4m_1B_{2} /T_{2} }<1-e^{-4m_1B_{1} /T_{1} }
\]
 hold in all the terms in Eq. (\ref{X1}), 
 since $m_1>0$. Similar to the uncoupled case,
 the sign of $X$ does not depend on the relation 
 between $P_{k}$ and $P_{k}^{'}$ for all $k=n/2+1,...,n$. 
 However, a definite relation
 between $P_{k}$ and $P_{k}^{'}$ for $k=2,3...,n/2$ 
 is not apparent under Eq. (\ref{lemma1}).
 
 Considering the BCS, mathematically written as,
\begin{equation}
 P_{k}^{'}>P_{k}; \quad k=2,3...,n/2  
 \label{PK1}
\end{equation}
  and then showing $X<0$ will prove 
  that $L_{X}<0$ or Eq. (\ref{lemma1}) 
  cannot make the coupled system work as an engine. 
  
  Now Eq. (\ref{PK1}) leads to the following conditions
\begin{equation}
P_{k}^{'}>P_{k}; \quad k=n/2+1,...,n-1  
    \label{PK2}
\end{equation}
For example using Eq. (\ref{lemma1}) 
and $P_{2}^{'}>P_{2}^{}$ (due to Eq. (\ref{PK1})) we have, 
\[
P_{n-1}^{'}=P_{2}^{'}. e^{-4(s-1)B_{2}/T_{2}} >
P_{n-1}^{}= P_{2}^{}. e^{-4(s-1)B_{1}/T_{1}}.
\] 
In this manner, all the relations given by
Eq. (\ref{PK2}) follow from
Eqs. (\ref{lemma1}) and (\ref{PK1}).
Also, as noted above, Eq. (\ref{lemma1}) implies 
$P_{n}^{'}>P_{n}^{}$.  
Therefore, using all these relations in 
the normalization condition of probabilities we obtain
\[
s\sum_{k=1}^{n/2}(P_{k}^{}-P_{k}^{'})>0.
\]
Relations (\ref{PK2}) along with $P_{n}^{'}>P_{n}^{}$
imply the following
\[
(s-r)(P_{n/2+1}-P_{n/2+1}^{'})<0,
...,(s-1)(P_{n-1}-P_{n-1}^{'})<0,s(P_{n}-P_{n}^{'})<0.
\]
Under BCS, we have
\[(-1).\left(P_{2}^{'}-P_{2} \right)<0,(-1).
\left(P_{3}^{'}-P_{3} \right)<0,...,(-r).
\left(P_{n/2}^{'}-P_{n/2} \right)<0.
\] 
Adding all the above inequalities, we obtain
the result that $X<0$.
This means that under Eq. (\ref{lemma1}), 
$L_{X}$ as well as $X$ are negative definite. \\

On the other hand, Eq. (\ref{cond}) implies
$Z_{2}>Z_{1}$, which further yields $P_{n}>P_{n}^{'}$. 
However, unlike the case with the uncoupled model, 
this does not determine the relative magnitudes
of the ground state probabilities $P_{1}$ and $P_{1}^{'}$  
(explicit expressions of these probabilities
are given in Table \ref{table:2}). 
Therefore, here we cannot be sure of the sign of 
the quantity $L_X$.  

Now, due to Eq. (\ref{cond}), we note that
\[
1-e^{-4m_1B_{2} /T_{2} }>1-e^{-4m_1B_{1} /T_{1} },
\] 
holds in all the terms in Eq. (\ref{X1}).
Also note that $X>0$ is always favored 
under this condition irrespective of the relation
between $P_{k}$ and $P_{k}^{'}$ for all $k=n/2+1,...,n$. 
The relation between $P_{k}$
and $P_{k}^{'}$ for $k=2,3...,n/2$ is also
not apparent under Eq. (\ref{cond}). As in the uncoupled case, 
here also we will consider the WCS, written as
\begin{equation}
 P_{k}^{'}<P_{k}; \quad k=2,3...,n/2.  
 \label{PK3}
\end{equation}
 Now, WCS leads to the following conditions
\begin{equation}
P_{k}^{'}<P_{k}; \quad k=n/2+1,...,n-1.  
    \label{PK4}
\end{equation} 
\begin{table}
\centering
\begin{tabular}{ p{10cm}} 
 \hline \hline
$P_{1}=e^{2sB_{1} /T_{1}} /Z_{1}$ \\ $P_{2} =e^{2(s-1)B_{1}/T_{1} +8sJ/T_{1}} 
/Z_{1}$ \\ $P_{3} =e^{2(s-1)B_{1} /T_{1}} /Z_{1}$ \\ $P_{4} 
=e^{2(s-2)B_{1} /T_{1} +8sJ/T_{1} +8J(s-1)/T_{1}} /Z_{1}$ \\ $P_{5}
=e^{2(s-2)B_{1} /T_{1} +8sJ/T_{1} } /Z_{1}$ \\ $P_{6} =e^{2(s-2)B_{1} /T_{1} }
/Z_{1}$ \\ ${.}$ \\ ${.}$ \\ ${.} $ \\ $P_{n/2-2s_{1}}
=e^{B_{1} /T_{1} +8sJ/T_{1} +8J(s-1)/T_{1} +8J(s-2)/T_{1} +...+
8J(s-\left(2s_{1} -1\right))/T_{1} } /Z_{1}$
\\  ${.}$ \\ ${.}$ \\ ${.}$ \\ 
$P_{n/2} =e^{B_{1} /T_{1}} /Z_{1}$ \\ $P_{n/2+1}
=e^{-B_{1} /T_{1} +8sJ/T_{1} +8J(s-1)/T_{1} +8J(s-2)/T_{1} +...+
8J(s-\left(2s_{1} -1\right))/T_{1} } /Z_{1}$ \\ ${.}$ \\
${.}$ \\ ${.}$ \\$P_{n/2+2s_{1} 
+1} =e^{-B_{1}/T_{1}} /Z_{1}$  \\ ${.}$ \\${.}$ \\ ${.}$
\\ $P_{n-5} =e^{-2(s-2)B_{1} /T_{1} +8sJ/T_{1} +8(s-1)J/T_{1}}
/Z_{1}$ \\ $P_{n-4} =e^{-2(s-2)B_{1} /T_{1} +8sJ/T_{1}} /Z_{1}$
\\ $P_{n-3} =e^{-2(s-2)B_{1}/T_{1}} /Z_{1}$ \\ $P_{n-2}
=e^{-2(s-1)B_{1}/T_{1}+8sJ/T_{1}} /Z_{1}$ \\ $P_{n-1} =e^{-
2(s-1)B_{1} /T_{1}} /Z_{1}$ \\  $ P_{n} =e^{-2sB_{1} /T_{1}} /Z_{1}$ \\
\hline \hline
\end{tabular}
\caption{\textit{Stage 1} occupation probabilities of the energy levels $E_{k}$
of the coupled spin system. $s_{1}
$ is smaller of the two spins in the terms involving the factor $2s_{1} +1$.}
\label{table:2}
\end{table}
For example, using Eq. (\ref{cond}) 
and $P_{2}>P_{2}^{'}$ (from (\ref{PK3})) 
we have
\[
P_{n-1}^{'}= P_{2}^{'}.e^{-4(s-1)B_{2}/T_{2}}<
P_{n-1}^{}= P_{2}^{}.e^{-4(s-1)B_{1}/T_{1}}
\] 
and thus, all the relations given by 
Eq. (\ref{PK4}) follow from Eqs. (\ref{cond}) 
and (\ref{PK3}). 
Also Eq. (\ref{cond}) implies $P_{n}>P_{n}^{'}$
as shown above. Thus, in total, we get
\begin{equation}
P_{k}^{'}<P_{k}^{}; \quad k=2,3,4,...,n.
    \label{WCS}
\end{equation}
Thereby, due to the normalization 
  condition on probabilities, we conclude
\begin{equation}
P_{1}^{'}>P_{1}.
    \label{P1P}
\end{equation}
In this manner, the WCS provides definite 
relations between the two probaility distributions.

We may combine Eqs. (\ref{WCS}) and (\ref{P1P}), to write 

\begin{equation}
\cfrac{P_{k}^{'} }{P_{1}^{'} }<\cfrac{P_{k} }{P_{1} }
    \implies \cfrac{e^{-E_{k}^{'}/T_{2}} }{e^{2sB_{2} /T_{2} } } 
    <\cfrac{e^{-E_{k} /T_{1}} }{e^{2sB_{1} /T_{1} } }, \quad k \neq 1. 
 \label{pkp1}
\end{equation}
Now, as shown in Appendix C, the above inequality yields  the
strictest condition on the permissible range of $J$, 
which is obtained for $k=2$, and is given as
\begin{equation}
0<J<\frac{\left(B_{2} -B_{1} \theta \right)}
{4s\left(1-\theta \right)} \equiv J_{c}.
\label{jc}
\end{equation} 
It implies that for $J$ to be in the above range,
all inequalities (\ref{pkp1}) hold good. 
\\
Now, using Eq. (\ref{PK4}) and $P_{n}>P_{n}^{'}$ 
in the normalization condition of probabilities we have,
\[ s\sum_{k=1}^{n/2}(P_{k}^{'}-P_{k})>0.
\]
Eq. (\ref{WCS}) implies the following
\[(s-r)(P_{n/2+1}-P_{n/2+1}^{'})>0,...
,(s-1)(P_{n-1}-P_{n-1}^{'})>0,s(P_{n}-P_{n}^{'})>0,\] 
\[(-1).\left(P_{2}^{'}-P_{2} \right),(-1).
\left(P_{3}^{'}-P_{3} \right),...,(-r).
\left(P_{n/2}^{'}-P_{n/2} \right)>0.
\] 
Adding all the above inequalities, we have $X>0$.
Therefore, we conclude that for WCS,
the following conditions ensure $X>0$:
$B_{2}>B_{1} \theta$ and $0<J<J_{c}$,
where  $\theta = T_2 /T_1$.

\noindent
Let us sum up the above discussion.
There are two relevant cases:
\par\noindent
a) $B_{2}<B_{1}\theta$, which implies the following:\\
1. $L_{X} \equiv (P_{1}^{'}-P_{1})+ (P_{n}-P_{n}^{'}) <0$.\\
2. $X<0$, thereby proving that
under 
$B_{2}<B_{1}\theta$, it is
not possible for the coupled system 
to work as an engine at all.
\par\noindent
b) $B_{2}>B_{1}\theta$, which  implies that:\\ 
1. $L_{X}$ does not bear a definite sign.
Although the term 
($P_{n}-P_{n}^{'}$) in $L_{X}$ 
is positive definite, yet the sign of the other term
($P_{1}^{'}-P_{1}$) is not definite.\\
2. Under WCS, we are able to prove $X>0$ for $B_{2}>B_{1} \theta$, 
thereby implying that it is a
necessary condition for $W_{av}>0$.
But, WCS also demands $P_{1}^{'}>P_{1}$
or $0<J<J_{c}$. Therefore,  
the latter constitutes a sufficient
condition for 
$W_{av}>0$.  \\
3.  
When $P_{1}^{'}>P_{1}$ does {\it not} hold,
$L_{X}$ does not have definite sign. 
So, depending on the control
 parameters, other terms in $X$ can be positive. 
 In this case, we cannot predict the sign of $W_{av}$.
 \\ We therefore conclude the following regarding 
 positive work extraction for the coupled model:\\
a. If, $L_X<0$  
(which happens for $B_{2}<B_{1}\theta$), then
$W_{av}<0$.\\ b. 
If $L_X >0$  
(which happens for $B_{2}>B_{1}\theta$ and $0<J<
J_{c}$), then $W_{av}>0$.
\\ c. If no definite sign can be assigned to $L_X$ 
(which may happen even when $B_{2}
>B_{1}\theta$ holds, but with no condition on the range of $J$), 
the system may or may not work as an engine.

\section{Condition on  \mathinhead{J}{J} from \mathinhead{W_{av}>0}{Wav}}
From the conditions given by Eqs.
(\ref{WCS}) and (\ref{P1P}), we obtained Eq. (\ref{pkp1}),
which leads to the following
\begin{equation}
\cfrac{E_{k}^{}}{T_{1}}-\cfrac{E_{k}^{'}}{T_{2}}
<2s\bigg(\cfrac{B_{2}}{T_{2}}-\cfrac{B_{1}}{T_{1}}\bigg).
\label{WCS2}
\end{equation} 
The above condition yields different
possible ranges for $J$ corresponding to different
energies $E_{k}$. Out of these, the shortest
range will clearly be permissible for 
{\it all} energy levels. Thus, we will
find the strictest condition on $J$ that
 ensures $W_{av}>0$.
Let us express
an energy eigenvalue as,  
\be
E_{k}^{}=m_{1} B_1-8m_{2} J, \quad 
E_{k}^{'}=m_{1} B_2-8m_{2} J.
\ee 
As can be seen from Fig. \ref{fig:1}, 
there are energy bands in the 
spectrum of the coupled system such 
that the energy levels corresponding to 
the same band have an identical value of $m_{1}$,
but different values of $m_{2}$. From the spectrum, 
we observe that 
$m_{1}$ varies from the minimum value of $-2s$ up to $2s$,
while $m_{2}$ can only take positive values (see Table \ref{table:5})
\begin{table}
\centering
\begin{center}
 \begin{tabular}{ p{2cm}| p{8cm} | p{4cm}} 
 \hline \hline
 $m_{1}$ & $m_{2}$ & $k$\\ 
 \hline \hline
$-2s$ & $0$ & $1$  \\ 
 \hline
$-2(s-1)$ & $s,0$ & $2,3$  \\
 \hline
 $-2(s-2)$ & $[s+(s-1)],s,0$ & $4,5,6$  \\
 \hline
 $-2(s-3)$ & $[s+(s-1)+(s-2)],[s+(s-1)],s,0$ & $7,..,10$\\
 \hline
 . & .&.\\
 \hline
  . & . &.\\
  \hline
  $-2(s-r)$ & $[s+(s-1)+...+(s-(2s_{1}-1))],...,0$ & $n/2-2s_{1},...,n/2$\\
 \hline 
 $2(s-r)$ & $[s+(s-1)+...+(s-(2s_{1}-1))],...,0$ & $n/2+1,...,n/2+2s_{1}+1$\\
 \hline
 . & .&.\\
 \hline
  . & . &.\\
  \hline
  $2(s-3)$ & $[s+(s-1)+(s-2)],[s+(s-1)],s,0$ & $(n-9),..,(n-6)$\\
  \hline
 $2(s-2)$ & $[s+(s-1)],s,0$ & $(n-5),(n-4),(n-3)$ \\
 \hline
 $2(s-1)$ & $s,0$ & $(n-2),(n-1)$  \\
 \hline
$2s$ & $0$ & $n$  \\ 
 \hline  \hline
 
\end{tabular}
\end{center}
\caption{ Spin dependent factors $m_{1}$ and $m_{2}$ when the energy eigenvalues
of the coupled system are expressed as: $E_{k}=m_{1}B-8m_{2}J$. The energy 
levels "$k$" which fall within the same band (i.e having same $m_{1}$) have also 
been specified.}
\label{table:5}
\end{table}
\noindent Eq. (\ref{WCS2}) now takes the following form 
\begin{equation}
    8m_{2}\bigg(\cfrac{1}{T_{2}}
    -\cfrac{1}{T_{1}}\bigg) J < (2s+m_{1})
    \bigg(\cfrac{B_{2}}{T_{2}}-
    \cfrac{B_{1}}{T_{1}}\bigg) 
    \implies J < \cfrac{(2s+m_{1})(B_{2}-B_{1}\theta)}{8m_2(1-\theta)}. 
    \label{COND J}
\end{equation}
Now within one
band (fixed value of $m_{1}$), it is obvious that the 
highest $m_{2}$ value will give the 
strictest condition on $J$.
Now, by referring 
to the spectrum, we infer
that for $m_{1} = -2(s - q)$, where $q=0,1,2,...$,  the 
largest value of $m_{2}$, denoted
by $m_{2,L}$ is \[m_{2,L}=s+ (s-1) +... + (s-q+1) 
= \frac{q}{2}(2s-q+1).\] Substituting
these values on R.H.S of Eq. (\ref{COND J}), 
we get the upper limit on $J$ as
\[
\cfrac{1}{2(2s-q+1)}
\cfrac{B_{2}-B_{1}\theta}{1-\theta}.
\]
Now, 
the strictest condition on the range of $J$
will be obtained for the lowest permissible
value of $q$, i.e. $q=1$ (since $m_2 =0$ for
$q=0$). Thus, we obtain
\begin{equation}
    0<J<\frac{1}{4s}.\cfrac{B_{2}
    -B_{1}\theta}{1-\theta} \equiv J_{c}.
    \label{jc2}
\end{equation}  
Therefore, we conclude that for the 
above range we have $W_{av}>0$. 
Note that Eq. (\ref{jc2}) is obtained
for $m_{1}=-2(s-1)$ and $m_{2}=s$,
which corresponds to the first 
excited state of the coupled system. 
\section{Proof for \mathinhead{X>Y_{1}}{X>Y1}}
\noindent As discussed in the main text,
for proving $\Delta S_{av}>0 $ we need to show,
\[X>Y_{1} \] 
\[\Rightarrow X-Y_{1} >0\Rightarrow 
X+(Y/s)>0\] for a case where all the 
terms of $Y_{}$ are negative.
\\ We will 
show that the PWCs, given by
Eqs (\ref{pwc}) and (\ref{jc0}),
derived for the coupled
model are enough to show the above
 relation and hence $\Delta S_{av}>0 $.
As already proved in the previous
sections that with $B_{2}>B_{1}\theta$,
the condition $J<J_{c}$ is obtained
by combining the following set of
conditions and then substituting $k=2$:
\begin{equation}
  P_{k}^{'}<P_{k}^{};\quad  k 
  \ge 2
\label{rell}
\end{equation} 
\[P_{1}^{'}>P_{1}\] 
 Eq. (\ref{rell}) also implies
 maximally negative $Y$. 
 $U \equiv X+Y/s>0$ will now be proved using Eq. (\ref{rell}) where $X$ 
 and $Y$ are given by Eq. (\ref{xy}). 
 \\ Before starting the proof, note that all the levels contribute to $X$ but
 only the $J$ dependent levels contribute to $Y/s$. The steps followed for 
proving $U>0$ under relations Eq. (\ref{rell}) are:\\ 1. We first consider the 
lower half levels. With $m_{1}$ being negative for all $k=1,..,n/2$ (see Table 
\ref{table:5}), the total contribution from these levels to $X$ takes the form,
\[ \cfrac{1}{2} \sum_{k=1}^{n/2} |m_{1}|(P_{k}^{'}-P_{k})\] 
\begin{table}
\centering
\begin{center}
 \begin{tabular}{p{5cm} p{8cm}} 
 \hline
 $k$ & $m_{3}=s+m_{4}$ \\ 
 \hline\hline
$1$ & $s$  \\ 
 \hline
$2,3$ & $s,(s-1)$  \\
 \hline
 $4,5,6$ & $(s-\frac{1}{s}),(s-1),(s-2)$  \\
 \hline
 $7,8,9,10$ &
 $(s-\frac{1}{s}-\frac{2}{s}),(s-1-\frac{1}{s}),(s-2),(s-3)$\\
 \hline
 . & .\\
  . & .\\
  \hline
  $(n/2-2s_{1}),...,n/2$ & $\left[s-(r-2s_{1})
-\frac{1}{s} -...-
\frac {\left(2s_{1} -1\right)}{s} \right],...,(s-r)$ \\
 \hline
 \hline
\end{tabular}
\end{center}
\caption{Coefficients $m_{3}$ of 
the terms ($P_{k}^{'}-P_{k}$) in $U$ with $k$ varying from $1,2,...,n/2$.}
\label{table:3}
\end{table} 

Similarly, the coefficients of these terms in $Y/s$ can be calculated from 
Table \ref{table:5} as $m_{2}/s$ (note that $m_{2}>0$ holds for all $k$). 
\\ Now we add these to get the coefficients of these terms in $U$, 
denoted by $m_{3} \equiv \cfrac{|m_{1}|}{2}+\cfrac{m_{2}}{s} $, which have 
been listed in Table \ref{table:3}. As can be seen, $m_{3}$ has a positive part 
given by "s" and a negative part, say $m_4$. The total contribution from the 
lower half levels  to $U$ is therefore written as, 
\[\sum_{k=1}^{n/2}\left(\cfrac{|m_{1}|}{2}+\cfrac{m_{2}}{s}\right)(P_{k}^{'}-P_{
k})=\sum_{k=1}^{n/2} m_{3}(P_{k}^{'}-P_{k})\] 
\[=\sum_{k=1}^{n/2}(s+m_{4})(P_{k}^{'}-P_{k}) 
=s\sum_{k=1}^{n/2}(P_{k}^{'}-P_{k})+\sum_{k=1}^{n/2} m_{4}(P_{k}^{'}-P_{k})\] 
With $m_{4}<0$, the second part is positive because of Eq. (\ref{rell}) and the 
first part is considered later on.\\ 2. We now consider the upper half levels. 
The total contribution of these levels to $X$ and $Y/s$ is considered 
separately. The former is given as, \[ \sum_{k=n/2+1}^{n} \cfrac{m_{1}}{2} 
(P_{k}^{}-P_{k}^{'})\] With $m_{1}$ being positive (Table \ref{table:5}) for all 
$k=n/2+1,..,n$, the above expression is positive because of Eq. (\ref{rell}).\\ 
As for these levels' contribution to $Y/s$, it is given 
as,\[\sum_{k=n/2+1}^{n-2} \cfrac{m_{2}}{s} (P_{k}^{'}-
P_{k}) \equiv \sum_{k=n/2+1}^{n-2} (m_{5}+m_{6})(P_{k}^{'}-
P_{k}^{}) =\sum_{k=n/2+1}^{n-2} m_{5}(P_{k}^{'}-
P_{k}^{})+\sum_{k=n/2+1}^{n-2} m_{6}(P_{k}^{'}-
P_{k}^{})\] Note that not all the levels contribute to $Y$ because many levels 
do not explicitly depend on $J$.
\begin{table}
\centering
\begin{center}
 \begin{tabular}{p{7cm} p{6cm} p{1.5 cm}} 
 \hline
 $k$ & $m_{2}/s=m_{5}+m_{6}$ & $m_{5}$\\ 
 \hline\hline
$(n/2+1),...,(n/2+2s_{1}+1)$ & $2s_{1}-\frac{1}{s} -...-
\frac {\left(2s_{1} -1\right)}{s},...,0$ & $2s_{1},...,0$  \\ 

 \hline
& . & .\\
 & . & .\\
$(n-9),(n-8),(n-7),(n-6)$ & $(3-\frac{1}{s}-\frac{2}{s}),(2-\frac{1}{s}),1,0$ 
& $3,2,1,0$ \\
 \hline
 $(n-5),(n-4),(n-3)$ & $(2-\frac{1}{s}),1,0$ & $2,1,0$ \\
 \hline
 $(n-2),(n-1)$ & $1,0$ & $1,0$\\
 \hline
 $n$ & $0$ & $0$ \\
 \hline
 \hline
\end{tabular}
\end{center}
\caption{Coefficients $m_{2}/s$
(obtained from Table \ref{table:5})
of the terms ($P_{k}^{'}-P_{k}$) in $Y/s$ with $k$ 
running over all upper half energy
levels i.e $k=n/2+1,...,n$.}
\label{table:4}
\end{table} 
Here $m_{5}$ and $m_{6}$ are respectively the positive and negative parts of 
$m_{2}/s$ (see Table \ref{table:4}). The second part in the above equation is 
positive because of (\ref{rell}) and the first part is considered later on.
\\3. 
Adding up the total contribution to $U$ from all the energy levels we have, 
\[\sum_{k=1}^{n/2}(s+m_{4})(P_{k}^{'}-P_{k})+\sum_{k=n/2+1}^{n} \cfrac{m_{1}}{2} 
(P_{k}^{}-P_{k}^{'})+\sum_{k=n/2+1}^{n-2} (m_{5}+m_{6})(P_{k}^{'}-
P_{k}^{})\] From the first and second points, we now have two parts which are 
yet to be proved positive. Their sum is given as, \begin{equation}
    \sum_{k=1}^{n/2}s(P_{k}^{'}-P_{k})+\sum_{k=n/2+1}^{n-2} m_{5}(P_{k}^{'}-
P_{k}^{})
\label{u}
\end{equation}
Using relations like $P_{n}^{'}<P_{n}$ and $P_{n-1}^{'}<P_{n-1}$ (from Eq. 
(\ref{rell})), and $P_{1}^{'}>P_{1}$ in the
normalization condition of 
probabilities 
\[\sum_{k=1}^{n}(P_{k}^{'}-P_{k})=0\]
 we have, \[\sum_{k=1}^{n/2}s(P_{k}^{'}-P_{k})+
\sum_{k=n/2+1}^{n-2}s(P_{k}^{'}-P_{k}^{})>0.
\] 
As shown below, $m_{5}<s$.
Therefore, with $P_{k}^{'}<P_{k}$,
we can safely replace $s$  by $m_{5}$ in the
above inequality, thereby proving $U>0$. 
  \subsection{Proof for  \mathinhead{m_{5}<s}{m5<s}}
To prove $m_{5}<s$, consider as an example, 
 the $k=n-5$ level, the explicit expression of the occupation probability is,
\[P_{n-5} =e^{-2(s-2)B_{1}
/T_{1} +8sJ/T_{1} +8J(s-1)/T_{1} } 
/Z_{1} \] and $m_{5}=2$ (see Table \ref{table:4}).
On carefully observing the energy spectrum, 
it can be seen that the energy level 
corresponding to this occupation probability
exists only if the sum of spins $"s"$ in the power
of the exponent satisfies, $2<s$. Similarly
$1<s$ holds in $P_{n-2}$, $3<s$ holds
in $P_{n-9},...,P_{n-6}$. \noindent
So this is true for
all the energy eigenvalues.
Since $\left(P_{n-5}^{'} -P_{n-5} \right)<0$ 
(Eq. (\ref{rell})) therefore we have 
 \[s\left(P_{n-5}^{'} -P_{n-5} \right)<
 2\left(P_{n-5}^{'} -P_{n-5} \right).\] 
\noindent Similarly we have the following:
\[s\left(P_{n-2}^{'} -P_{n-2} \right)
<1\left(P_{n-2}^{'} -P_{n-2} \right),\]
\[s\left(P_{n-4}^{'} -P_{n-4} \right)
<1\left(P_{n-4}^{'} -P_{n-4} \right),\] 
\[\vdots\]
\[s\left(P_{n/2+1}^{'} -P_{n/2+1}
\right)<2s_{1}\left(P_{n/2+1}^{'} -P_{n/2+1} \right).\] 
\noindent The last inequality
follows from the fact $s_{1}<s_{2}$.
This proves  $m_{5}<s$.
\begin{center}
   \textbf{Case study: $\mathbf{s_{1}=1/2, s_{2}=1}$ } 
\end{center}
\noindent As an illustration of the above proof for 
the upper bound of Otto efficiency, we consider the ($1/2, 1$)
coupled system (Fig. \ref{fig:8}), where $n=6$.
\begin{figure}
\centering
\begin{tikzpicture}
\scalebox{10}{5}
\draw (2cm,0em) -- (8cm,0em) node[right]
{$E_{1}=-2sB$};
\draw (2cm,2.5em) -- (8cm,2.5em) node[right]
{$E_{2}=-2(s-1)B-8sJ$};
\draw (2cm,4.5em) -- (8cm,4.5em) node[right]
{$E_{3}=-2(s-1)B$};
\draw (2cm,6.5em) -- (8cm,6.5em) node[right]
{$E_{4}=2(s-1)B-8sJ$};
\draw (2cm,9em) -- (8cm,9em) node[right]
{$E_{5}=2(s-1)B$};
\draw (2cm,11.7em) -- (8cm,11.7em) node[right]
{$E_{6}=2sB$};
\end{tikzpicture} \\
\caption{Energy levels $E_{k}$ of the coupled
two-spins system $(1/2, 1)$, where $s=3/2$.}
\label{fig:8}
\end{figure}
\noindent The \textit{Stage 1} equilibrium occupation probabilities of these 
levels are of the form:
$P_{k} = {e^{-m_{1} B_{1}/T_{1} +8m_{2} J/T_{1}}}/{Z_{1}}$,
 where the spin-dependent factors $m_{1}$ and $m_{2}$ have been 
 specified in Table \ref{table:6} and the  partition function is
\[
Z_{1}= \begin{array}{l} {\cal Z}_{1}+2\cosh 
\left[2(s-1)B_{1}/T_{1}\right].e^{8sJ/T_{1}}
\end{array}
\]
with
 \[
 {\cal Z}_{1} \equiv 2\sum_{k=1}^{s+1/2}
\cosh{\left[2(s-k+1)B_{1} /T_{1}\right]}=2 \left(\cosh[2sB_1/T_1]+
\cosh[2(s-1)B_1/T_1] \right).
\]  
\begin{table}
\centering
\begin{center}
 \begin{tabular}{ p{2cm}| p{2cm} | p{2cm}} 
 \hline \hline
 $m_{1}$ & $m_{2}$ & $k$\\ 
 \hline \hline
$-2s$ & $0$ & $1$  \\ 
 \hline
$-2(s-1)$ & $s,0$ & $2,3$  \\
 \hline
 $2(s-1)$ & $s,0$ & $4,5$  \\
 \hline
$2s$ & $0$ & $6$  \\ 
 \hline  \hline
 
\end{tabular}
\end{center}
\caption{}
\label{table:6}
\end{table}
The heat absorbed from the hot bath and average work are given as, 
\[Q_{1,av}=2B_{1}X+8JY, \quad W_{av}=2(B_{1}-B_{2})X, \] 
where
\[
X= \cfrac{1}{2}\sum_{k=1}^{n=6} m_{1} (P_{k}^{}-
P_{k}^{'}), \quad Y=  \sum_{k=2}^{(n-2)=4} 
m_{2} (P_{k}^{'}-P_{k})=(P_{2}^{'}-
P_{2}^{})+(P_{4}^{'}-
P_{4}^{}).\]

\begin{center}
    \textbf{Proof for $\mathbf{X>Y_{1}}$}
\end{center}
\noindent As discussed in the main text, we will be proving $U=X+(Y/s)>0$ 
using Eq. (\ref{rell}) and the 
condition $P_{1}^{'}>P_{1}$.
Note that all levels contribute to $X$ but only the $J$ dependent levels 
($E_{2}$ and $E_{4}$) contribute to 
$Y$. The steps followed for proving $U>0$ under relations Eq. (\ref{rell}) 
are:
\\ 1. We first consider the lower half ($k=1,2,3$) of the levels. With $m_{1}$
being negative (see Table \ref{table:6}), the total contribution from these 
levels to $X$ takes the form,
\[ \cfrac{1}{2} \sum_{k=1}^{3} |m_{1}|(P_{k}^{'}-P_{k})=s(P_{1}^{'}-P_{1})
+(s-1)(P_{2}^{'}-P_{2}+P_{3}^{'}-P_{3})\] 
\noindent Similarly contribution of lower half levels to $Y/s$ is written as,
\[Y/s=\cfrac{m_{2}}{s}(P_{2}^{'}-P_{2})=(P_{2}^{'}-P_{2})\] Total contribution 
of lower half levels to $U$ is, 
\[U=X+Y/s=s(P_{1}^{'}-P_{1})+(s-1)(P_{2}^{'}-P_{2}+P_{3}^{'}-P_{3})+(P_{2}^{'}
-P_{2})\]

\[U=s(P_{1}^{'}-P_{1})+s(P_{2}^{'}-P_{2})+(s-1)(P_{3}^{'}-P_{3})=
\sum_{k=1}^{3} s(P_{k}^{'}-P_{k})+(-1)(P_{3}^{'}-P_{3})\]
The second part is positive because of Eq. (\ref{rell}) and the first part is 
considered later on.\\ 2. We now consider the upper half levels. The total 
contribution of these levels to $X$ and $Y/s$ is considered separately. The 
former is given as, \[ \sum_{k=4}^{6} \cfrac{m_{1}}{2} 
(P_{k}^{}-P_{k}^{'})=(s-1)(P_{4}-P_{4}^{'}+P_{5}-P_{5}^{'})+s(P_{6}^{}-P_{6}^{'}
)\] The above expression is positive because of Eq. (\ref{rell}).\\ As for these 
levels' contribution to $Y/s$, it is given as,
\[Y/s=\cfrac{m_{2}}{s} (P_{4}^{'}-
P_{4})=(P_{4}^{'}-P_{4})\]
This part is negative and will be considered later on. \\ 3. From points 1. 
and 2., the following terms in $U$ are yet to be shown positive,
\[\sum_{k=1}^{3} s(P_{k}^{'}-P_{k})+(P_{4}^{'}-P_{4})\]
Using relations like $P_{6}^{'}<P_{6}$, $P_{5}^{'}<P_{5}$ 
(from Eq. (\ref{rell})) and $P_{1}^{'}>P_{1}$ in the
normalization condition of 
probabilities, given as, 
\[\sum_{k=1}^{6}(P_{k}^{'}-P_{k})=0\]
 we have, \[\sum_{k=1}^{4}s(P_{k}^{'}-P_{k})>0 \implies 
 \sum_{k=1}^{3}s(P_{k}^{'}-P_{k})+s(P_{4}^{'}-P_{4})>0 \] 
With $P_{4}^{'}<P_{4}$ and $1<s$ ($s=3/2$ for the present case), we can 
safely replace $s$  by $1$ in the
the above expression thereby proving $U>0$. 
\bibliography{references.bib}      
\end{document}